\documentclass{aa520}
\usepackage{psfig,graphics}
\usepackage{amssymb}
\usepackage{txfonts}  % Hugh, please use the v5.x A&A style file - Nigel
\DeclareMathAlphabet{\scriptnew}{U}{eus}{m}{n}
\begin{document}
\bibliographystyle{plain}
%
%\setlength{\parindent}{3em}
%\thesaurus{02         % A&A Section 12: Atomic, molecular and nuclear data
%               (02.01.4;  % Atomic processes
%               02.03.2)} % Atomic data
%
\title{Dielectronic recombination data for dynamic finite-density plasmas}
\subtitle{I. Goals and methodology}
\author{N.~R.~Badnell
\inst{1}
\and M.~G.~O'Mullane
\inst{1}
\and H.~P.~Summers
\inst{1}
\and Z.~Altun
\inst{2}
\and M.~A.~Bautista
\inst{3}
\and J.~Colgan
\inst{4}
\and T.~W.~Gorczyca
\inst{5}
\and D.~M.~Mitnik
\inst{6}
\and M.~S.~Pindzola
\inst{4}
\and O.~Zatsarinny
\inst{5}
}
\institute{Department of Physics, University of Strathclyde, Glasgow G4 0NG, UK
\and Department of Physics, Marmara University, 81040, Ziverbey, Istanbul,  Turkey
\and Centro de F\'{\i}sica, Instituto Venezolano de Investigaciones 
Cient\'{\i}ficas (IVIC), PO Box 21827, Caracas 1020A, Venezuela
\and Department of Physics, Auburn University, Auburn, AL 36849, USA
\and Department of Physics, Western Michigan University, Kalamazoo, MI 49008, USA
\and Departamento de F\'{\i}sica, FCEN, Universidad de Buenos Aires, Buenos Aires, Argentina
}   
\offprints{N. R. Badnell}
\date{Received \today}
\abstract{
A programme is outlined for the assembly of a comprehensive dielectronic
recombination database within the generalized collisional--radiative (GCR)
framework.  It is valid for modelling ions of elements in dynamic finite-density
plasmas such as occur in transient astrophysical plasmas such as solar flares 
and in the divertors and high transport regions of magnetic fusion devices. The
resolution and precision of the data are tuned to spectral analysis and so are
sufficient for prediction of the dielectronic recombination contributions to
individual spectral line emissivities.  The fundamental data are structured
according to the format prescriptions of the Atomic Data and Analysis Structure
(ADAS) and the production of relevant GCR derived data for application is
described and implemented following ADAS.  The requirements on the dielectronic
recombination database are reviewed and the new data are placed in context and evaluated
with respect to older and more approximate treatments. Illustrative results
validate the new high-resolution zero-density dielectronic recombination data in
comparison with  measurements made in heavy-ion storage rings utilizing an
electron cooler. We also exemplify the role of the dielectronic data on GCR
coefficient behaviour for some representative light and medium weight elements.
}
\maketitle
%
%\titlerunning{Dielectronic recombination data for dynamic finite-density plasmas I.}
%\authorrunning{N.R. Badnell  et al.}
%________________________________________________________________

\section{Introduction}
\label{introduction} Dielectronic recombination (DR) is the dominant electron--ion
recombination process in many astrophysical and laboratory plasmas. It plays an
important role in  determining both the level populations and the ionization balance of 
both high- and  low-temperature non-LTE plasmas over a wide range of electron densities.
Dielectronic recombination can be viewed as a two-step process. Firstly, a free
electron excites an electron in an ion $X^{+z}$, say, and in the process transfers 
sufficient of its  energy that it is captured into an autoionizing state of the ion
$X^{+z-1}$. The process is reversible since the total energy of the system remains
conserved. However, if an electron (either the captured electron or an electron in
the parent core) then makes a spontaneous radiative transition that leaves the ion in a
non-autoionizing (bound) state then the recombination can be viewed as complete, at
least in the low-density (coronal) limit. Dielectronic recombination can take place
via many intermediate autoionizing states -- indeed, entire Rydberg series.  It was
the significance of the effective statistical weight of so many available states that
led Burgess (1964\cite{bur64}) to recognize the importance of dielectronic recombination in the
first instance. A consequence of this  is that a huge population structure model is
required in principle for further progress.  In astrophysics and fusion, the problem
has usually been made manageable by simply summing over all final states so as to produce 
a total dielectronic  recombination rate coefficient. In combination with radiative
recombination, this is the effective recombination coefficient from the point of view of 
an ionization state only -- the so-called `coronal picture'. In the latter, excited-state
populations are depleted exclusively by spontaneous radiative transitions and
are small compared to those of ground states -- with which they are in quasi-static
equilibrium. Collisional processes are negligible, except with ground state targets. 
In turn, the ionization state is determined by balancing the  effective recombination 
rate of (the ground state of) $X^{+z}$ against the collisional ionization rate of
(the ground state of) $X^{+z-1}$. 

For dynamic finite-density plasmas, there are therefore two critical limitations to the coronal
approximation which must be addressed.  Both require a re-appraisal of the strategy for computing
dielectronic recombination, which is the objective of this paper.  Firstly, from the atomic point of view, a
dynamic plasma is one for which the timescale of change in plasma parameters (especially electron
temperature, $T_{\rm e}$, and electron density, $N_{\rm e}$)  is comparable with the lifetime of metastable
populations of its constituent ions.  This situation is a concern, for example, for impurities near plasma
contacted surfaces in fusion (Summers  et al. 2002\cite{sum02}) and for active solar events (Lanza  et al.
2001\cite{lan01}). The
population of an ionization stage may no longer be assumed to be concentrated in the ground state of each
ion. A significant population is found in the metastables, and they are not in quasi-static equilibrium with
the ground state.  This means that such metastables may be the starting point for recombination events, and
so the time-evolution of their populations must be tracked in the same manner as for the ground states. We
call this the generalized collisional--radiative (GCR) picture (Summers \& Hooper 1983\cite{sum83}).

The second issue is that of finite-density effects. The coronal (zero-density limit)
picture assumes that, following the two-step dielectronic recombination process, the
resultant (non-autoionizing) excited-state (electron) radiatively cascades back down to the ground
state without collisional disruption. At finite electron densities, this radiative cascade
can be interrupted by further electron collisions which redistribute the population -- in
particular, ionization (possibly stepwise) out of excited states, which reduces the
effective dielectronic recombination rate. Collisional--radiative modelling  removes the
limitations of the coronal model, but at the cost of much more elaborate excited population
calculations.  These in turn require much more detailed dielectronic recombination data,
and in an easily accessible form. The onset of these density effects on the `post-DR'
population structure depends markedly on ion charge, but can be significant even at
electron densities as low as $N_{\rm e} \gtrsim 10^8 {\rm cm}^{-3}$ typical of the solar corona.  At much
higher densities, $N_{\rm e} \sim 10^{14} {\rm cm}^{-3}$, redistributive collisions can interrupt the
two-step dielectronic process itself.  

Therefore, we require dielectronic recombination from the metastable states as well
as the ground. Secondly, we require  final-state resolved data, i.e. we need to know
the specific level that each two-step recombination ends-up in -- the generalized
collisional--radiative population rate equations govern the subsequent time-evolution of
these states. (Incidentally, the first requirement imposes an additional requirement
on the second, namely, that we now require dielectronic recombination into metastable
parent final-states that lie above the ionization limit so that the
collisional--radiative modelling  recovers the Saha--Boltzman populations and
ionization fractions in the high-density  limit.) As we have already indicated, there
are very many possible final states accessible to the dielectronic recombination
process which would seem to make for unacceptably large tabulations. However, it is
not only impracticable, but unnecessary, to treat each final state in the same manner
for the purposes of collisional--radiative modelling. The techniques of matrix
condensation and projection reduce the effective number of high-lying states by the
progressive bundling of representative states and project the full influence of the
high-lying states down onto a fully-resolved low-lying set. (It is the low-level set
which is the focus of detailed spectral analysis.) With this in mind, we can
tailor our dielectronic recombination tabulations to reflect this situation.

Of course, many people have calculated both partial and total dielectronic
recombination rate coefficients and it is impractical to list them all here.  A useful
starting point is the compilation of total recombination rate coefficients from the
literature by  Mazzotta  et al. (1998\cite{maz98}) who then used this data to compute the coronal
ionization balance for all elements up to Ni. When the results of {\it ab initio}
calculations are not available then much use, and abuse, is made of the General
Formula of Burgess (1965\cite{bur65}). Partial dielectronic recombination rate coefficients also
abound in the literature in connection with the study of particular physical problems:
for example, at low temperatures, where only a few autoionizing states contribute
significantly (Nussbaumer \& Storey 1983\cite{nus83}); satellite lines (e.g. Bely-Dubau  et al.
1979\cite{bel79}); laser produced plasmas, where the electron density and/or charge state is high
enough that the non-LTE populations are concentrated in a limited number of low-lying
states (see e.g. Abdallah \& Clark 1994\cite{abd94}).

We have found it helpful for application to prepare and handle dielectronic recombination
data in a hierarchy of increasing sophistication which we call {\it baseline}, {\it level 1} and 
{\it level 2}. {\it Baseline} data are those produced using the Burgess general formula (GF),
or with the techniques and state-selective programs associated with it.  The generality of
dielectronic data in use today are still from the GF.  The background codes to the GF have a
capability significantly beyond that of the GF, but are less well known.  We return to these in more
detail in section \ref{burgess-bethe}. The {\it level 1} approach was introduced in the early `ninties'
to support the GCR modelling of light elements such as Be, B and C.  These species are used in
`light-element strategies' for plasma facing wall components in fusion technology.  For
such elements, it is sufficient to use LS-coupled atomic structure and term population
modelling. The relevant metastable populations in this case are terms such as
C$^{2+}(2{\rm s}2{\rm p}~^3{\rm P})$. The {\it level 2} approach, which is the main purpose of the present
paper series, is concerned with the need to handle medium and heavy species in both fusion and
astrophysics and to handle more extreme environments. In this approach, we aim to work with levels
rather than terms and to use an intermediate coupling scheme, based-on the use of the
Breit--Pauli Hamiltonian (Badnell 1997\cite{bad97}).  

There are several reasons why the {\it level 2} approach is now necessary: astrophysical spectral
diagnostics tend to be based on levels rather than terms and the competition between
autoionization and radiation makes it difficult to partition term-resolved
dielectronic recombination data over levels; nuclear spin--orbit mixing is important
even for low-$Z$ ions, e.g. carbon, because only a weak mixing of LS-forbidden
autoionization rates with LS-allowed can give rise to `forbidden' autoionization
rates that are comparable with the dominant radiative rates (see Nussbaumer \&
Storey 1984\cite{nus84}, Badnell 1988\cite{bad88}); as $Z$ increases further then LS-forbidden radiative
rates start to  become significant; finally, dielectronic recombination via
fine-structure transitions is completely absent in LS-coupling, giving rise to a large
underestimate of the low-temperature dielectronic recombination rate coefficient  in
some iso-electronic sequences (see Savin  et al. 1997\cite{sav97}).

The goal of this work is to calculate multi-configuration intermediate coupling dielectronic recombination 
rate coefficients from the (ground plus) metastable levels of an ion to all possible final states,
resolved by level, and/or bundling, appropriate for generalized collisional--radiative modelling.
We will cover elements applicable to astrophysics and magnetic fusion viz. He, Li, Be, B, C,
N, O, F, Ne, Na, Mg, Al, Si, P, S, Cl, Ar, Ca, Ti, Cr, Fe, Ni, Zn, Kr, Mo and Xe. The first phase
of the work will be the H- through Ne-like sequences. {\it Level 1} LS-coupling data for
many elements of these sequences was calculated by Badnell (1991--92, unpublished) and
incorporated into the atomic database part of the Atomic Data and Analysis Structure (ADAS)
and is routinely  drawn into the generalized collisional--radiative part of ADAS (see
Summers 2001). So, we already have a clear pathway through to the complete utilization  of
the detailed {\it level 2} data that we will produce. The second phase of the work will
cover the Na-like through Ar-like sequences, piloted initially by further ({\it level 1}) 
LS-coupling calculations to extend the 1991--92 data.  Note that ADAS uses year numbers for the
introduction of new approximations bringing substantive contributions to the database.   The
LS-coupled work of Badnell above has the year number '93'.  The third phase will focus on
the remaining sequences  of particular elements of interest, e.g. Fe. Even with the
compactification of the partial dielectronic recombination data, along the lines already
indicated, full publication in a paper journal is impractical, and not especially useful. 
So, the entire data will be made available via the World Wide Web (see section
\ref{database}).  The organization of the dielectronic data product follows the data format
specifications of the ADAS Project.  Dielectronic data are assigned to the data format {\it
adf09} (Summers 2001\cite{sum01}).

The plan for the remainder of the paper is as follows: in section \ref{gcr-modelling}
we review the generalized collisional--radiative approach encapsulated in ADAS and
which influences our approach to handling dielectronic  recombination data. In
section \ref{dr-coefft-modelling} we describe and justify the theoretical approach
that we take to calculate {\it level 2} data. In section \ref{burgess-bethe} we
review in some detail the essence of the Burgess approach.  It will be shown that
this remains of importance for a full exploitation of the new work, for example, as
applied to the $l$-redistribution of autoionizing states.  Also, it is
necessary to assess the progress in the precision of new dielectronic data in
comparison with the {\it baseline} data. In section \ref{experimental-validation} we
discuss the experimental situation for verifying dielectronic recombination data and the
role of external fields. Some comparisons of results of our theoretical approach with
high-resolution experimental results from storage rings are given.  In section
\ref{derived-data-validation} we address derived data and we present some
illustrative comparisons of GCR effective coefficients obtained using {\it baseline},
{\it level 1} and {\it level 2}  data from dielectronic calculations. Also, we illustrate 
metastable-resolved ionization fractions. In section \ref{database} we give more
detail of the organization of the database and the computational implementation of its
production.  We finish with a short summary.

\section{Theory}
\label{theory}

\subsection{Generalized collisional--radiative modelling}
\label{gcr-modelling}

Consider ions $X^{+z}$ of element $X$ of charge state $z$. We separate the levels
of $X^{+z}$ into metastable levels $X^{+z}_{\rho}$, indexed by Greek indices,
and excited levels, indexed by Roman indices. The metastable levels include
the ground level. We assume that the excited levels $X^{+z}_{i}$ are populated
by excitation from all levels, $\rho$ and $j$, of $X^{+z}$, by ionization from the 
metastable levels of $X^{+z-1}_{\mu}$ and recombination from the metastable
levels of $X^{+z+1}_{\nu}$. The dominant population densities of these ions
in the plasma are denoted by $N^{z}_{\rho}$, $N^{z-1}_{\mu}$ and $N^{z+1}_{\nu}$.
The excited-state population densities, $N^{z}_{i}$, are assumed to be in
quasi-static equilibrium with respect to the metastable populations. Thus,
\begin{eqnarray}
0&=&\sum_{\sigma}C^{z}_{i\sigma}N^{z}_{\sigma}+\sum_{j}C^{z}_{ij}N^{z}_{j}
+\sum_{\mu}S^{z-1}_{i\mu}N^{z-1}_{\mu} 
%\nonumber \\
+\sum_{\nu}R^{z+1}_{i\nu}N^{z+1}_{\nu},
\label{eq1}
\end{eqnarray}
where $C^{z}_{ij}$ are elements of the collisional--radiative matrix defined
by 
\begin{eqnarray}
N_{{\rm e}}C^{z}_{ij}=A^{{\rm r}}_{j\rightarrow i}+N_{{\rm e}}q^{\rm e}_{j\rightarrow i}\,,
\hspace{10mm}j>i\,,
\label{eq2}
\end{eqnarray}
where $N_{{\rm e}}$ is the electron density, $q^{\rm e}_{j\rightarrow i}$ is the
electron-impact de-excitation rate coefficient and $A^{{\rm r}}_{j\rightarrow i}$ is the
spontaneous radiative rate, both for the $j\rightarrow i$ transition in the ion $X^{+z}$.
The equivalent expressions for upward transitions ($j<i$) and for $i$ and/or $j$
replaced by a metastable index follow trivially. The diagonal element $C^{z}_{ii}$
denotes the loss rate coefficient from the excited state $i$ and is given by
\begin{eqnarray}
C^{z}_{ii}=-\sum_{j\neq i}C^{z}_{ji}-S^{z}_{i}\,,
\label{eq3}
\end{eqnarray}
where
\begin{eqnarray}
S^{z}_{i}=\sum_{j}S^{z}_{ji}
\label{eq4}
\end{eqnarray}
is the total ionization rate coefficient out of $i$.
Finally, $S^{z-1}_{i\mu}$ is the partial ionization rate coefficient out of
metastable $\mu$ and $R^{z+1}_{i\nu}$ is the partial
recombination rate coefficient out of metastable level $\nu$, both
into level $i$ of the ion $X^{+z}$. $S^{z-1}_{i\mu}$ includes contributions
from both direct ionization and excitation-autoionization. $R^{z+1}_{i\nu}$
includes contributions from three-body, radiative and dielectronic
recombination.

Solving for $N^{z}_{j}$, we have
\begin{eqnarray}
N^{z}_{j}&=&-\sum_{\sigma,i}\left(C^{z}\right)^{-1}_{ji}C^{z}_{i\sigma}N^{z}_{\sigma}
-\sum_{\mu,i}\left(C^{z}\right)^{-1}_{ji}S^{z-1}_{i\mu}N^{z-1}_{\mu} \nonumber \\
&&-\sum_{\nu,i}\left(C^{z}\right)^{-1}_{ji}R^{z+1}_{i\nu}N^{z+1}_{\nu} \nonumber \\
&\equiv&\sum_{\sigma}{N_{\rm e}\, ^{{\rm X}}{\cal F}^{z}_{j\sigma}N^{z}_{\sigma}}
+\sum_{\mu}{N_{\rm e}\, ^{{\rm I}}{\cal F}^{z}_{j\mu}N^{z-1}_{\mu}}
+\sum_{\nu}{N_{\rm e}\, ^{{\rm R}}{\cal F}^{z}_{j\nu}N^{z+1}_{\nu}} \nonumber \\
&\equiv&\sum_{\sigma}{^{{\rm X}}\!N^{z}_{j\sigma}}
+\sum_{\mu}{^{{\rm I}}\!N^{z}_{j\mu}}
+\sum_{\nu}{^{{\rm R}}\!N^{z}_{j\nu}}\, ,
\label{eq51}
\end{eqnarray}
where $^{{\rm X}}\!N^{z}_{j\sigma}$, $^{{\rm I}}\!N^{z}_{j\mu}$ and $^{{\rm R}}\!N^{z}_{j\nu}$
are the effective populations of $j$ due to excitation, ionization and
recombination from their respective metastables, and $^{{\rm X}, {\rm I}, {\rm R}}{\cal F}^z_{j\beta}$
the corresponding coefficients.
It is here that the connection with spectral analysis is made.
The total emissivity in the line $j\rightarrow k$ is given by
\begin{eqnarray}
\varepsilon^{z}_{j\rightarrow k}&=&N^{z}_{j}A^{\rm r}_{j\rightarrow k}\, ,
\label{eq52}
\end{eqnarray}
with the populations $N^{z}_{j}$ given by equation (\ref{eq51}). The corresponding photon
emissivity coefficients are defined by
\begin{eqnarray}
^{{\rm X}, {\rm I}, {\rm R}}\scriptnew{PEC}^{z}_{\beta, j\rightarrow k}&\equiv&^{{\rm X}, 
{\rm I}, {\rm R}}{\cal F}^z_{j\beta}
A^{\rm r}_{j\rightarrow k}\, ,
\label{eq53}
\end{eqnarray}
respectively.
Thus, the contributions to the total $j\rightarrow k$ emissivity from excitation, ionization and
recombination are given by
\begin{eqnarray}
^{{\rm X}, {\rm I}, {\rm R}}\varepsilon^{z}_{j\rightarrow k}&=&N_{\rm e} \sum_\beta
{ ^{{\rm X}, {\rm I}, {\rm R}}
\scriptnew{PEC}^{z}_{\beta, j\rightarrow k}N^{z,\, z-1,\, z+1}_\beta}\, ,
\label{eq54}
\end{eqnarray}
respectively.

The dynamic metastable populations $N^{z}_\rho$ of $X^{z}$ satisfy
\begin{eqnarray}
\frac{1}{N_{{\rm e}}}\frac{{\rm d} N^{z}_{\rho}}{{\rm d} t}&=&
\sum_{\sigma}\left\{C^{z}_{\rho\sigma}N^{z}_{\sigma}+
\sum_{j}C^{z}_{\rho j}\,{^{{\rm X}}\!N^{z}_{j\sigma}}\right\} \nonumber \\
 &-&\sum_{\sigma}\left\{C^{z}_{\sigma\rho}N^{z}_{\rho}+
\sum_{j}C^{z}_{\sigma j}\,{^{{\rm X}}\!N^{z}_{j\rho}}\right\} \nonumber \\
&+&\sum_{\mu}\left\{S^{z-1}_{\rho\mu}N^{z-1}_{\mu}+
\sum_{m}S^{z-1}_{\rho m}\,{^{{\rm X}}\!N^{z-1}_{m\rho}}\right\} \nonumber \\
 &-&\sum_{\nu}\left\{S^{z}_{\nu\rho}N^{z}_{\rho}+
\sum_{j}S^{z}_{\nu j}\,{^{{\rm X}}\!N^{z}_{j\rho}}\right\} \nonumber \\
&+&\sum_{\nu}\left\{R^{z+1}_{\rho\nu}N^{z+1}_{\nu}+
\sum_{j}C^{z}_{\rho j}\,{^{{\rm R}}\!N^{z}_{j\nu}}\right\} \nonumber \\
 &-&\sum_{\mu}\left\{R^{z}_{\mu\rho}N^{z}_{\rho}+
\sum_{m}C^{z-1}_{\mu m}\,{^{{\rm R}}\!N^{z-1}_{m\rho}}\right\} \nonumber \\
&+&\sum_{\sigma}\left\{Q^{z}_{\rho\sigma}N^{z}_{\sigma}+
\sum_{m}S^{z-1}_{\rho m}\,{^{{\rm R}}\!N^{z-1}_{m\sigma}}\right\} \nonumber \\
 &-&\sum_{\sigma}\left\{Q^{z}_{\sigma\rho}N^{z}_{\rho}+
\sum_{m}S^{z-1}_{\sigma m}\,{^{{\rm R}}\!N^{z-1}_{m\rho}}\right\} \nonumber \\
&\equiv&\sum_{\sigma}X^{z\rightarrow z}_{{\rm CD}:\sigma\rightarrow\rho}N^{z}_{\sigma}
-\sum_{\sigma}X^{z\rightarrow z}_{{\rm CD}:\rho\rightarrow\sigma}N^{z}_{\rho} \nonumber \\
&+&\sum_{\mu}S^{z-1\rightarrow z}_{{\rm CD}:\mu\rightarrow\rho}N^{z-1}_{\mu}
-\sum_{\nu}S^{z\rightarrow z+1}_{{\rm CD}:\rho\rightarrow\nu}N^{z}_{\rho} \nonumber \\
&+&\sum_{\nu}\alpha^{z+1\rightarrow z}_{{\rm CD}:\nu\rightarrow\rho}N^{z+1}_{\nu}
-\sum_{\mu}\alpha^{z\rightarrow z-1}_{{\rm CD}:\rho\rightarrow\mu}N^{z}_{\rho} \nonumber \\
&+&\sum_{\mu}Q^{z\rightarrow z}_{{\rm CD}:\sigma\rightarrow\rho}N^{z}_{\sigma}
-\sum_{\sigma}Q^{z\rightarrow z}_{{\rm CD}:\rho\rightarrow\sigma}N^{z}_{\rho}\,,
\label{eq6}
\end{eqnarray}
which defines the generalized collisional--radiative excitation ($X_{\rm CD}$), ionization
($S_{\rm CD}$), recombination ($R_{\rm CD}$) and parent metastable cross-coupling ($Q_{\rm CD}$)
rate coefficients. (We note that $Q^{z}_{\rho\sigma}\equiv 0$
initially.) This set of equations, together with those for $N^{z}_{j}$, are sufficient
to solve the low-level problem -- those levels with principal quantum number $n\le n_{{\rm c}}$, say.
In the absence of dielectronic recombination, or at sufficiently high electron densities
(e.g. $\gtrsim10^{18} {\rm c}{\rm m}^{-3}$, say, as found in laser-produced plasmas) then $n_{{\rm c}}$
can be small enough for this to be a complete solution since all higher levels are in
collisional LTE, with a Boltzman population distribution. However, the presence of
dielectronic recombination in low- to medium-density plasmas means that $n_{{\rm c}}$ can
be prohibitively large ($\approx 500$, say). 

We now consider a projection-condensation
approach that allows for the effect of the high-level populations ($\overline{n}\equiv n>n_{{\rm c}}$)
on the low-level populations ($n\le n_{{\rm c}}$). We work in the bundled-$n$ picture.
Here the populations are grouped according to their parent level and principal quantum number.
We assume that the high-level populations are in quasi-static equilibrium with the low-level
populations and adjacent stage metastables. Thus, for each parent $\tau$, the high-level
populations (denoted by $\tau\overline{n}$) satisfy
\begin{eqnarray}
0&=&\sum_{n}C^{z}_{\tau\overline{n},\tau n}N^{z}_{\tau n}
+\sum_{\overline{n}^{\prime}}C^{z}_{\tau\overline{n},\tau\overline{n}^{\prime}}
N^{z}_{\tau\overline{n}^{\prime}} \nonumber \\
&+&\sum_{\mu}S^{z-1}_{\tau\overline{n},\mu}N^{z-1}_{\mu}
+\sum_{\nu}R^{z+1}_{\tau\overline{n},\nu}N^{z+1}_{\nu}\, ,
\label{eq7}
\end{eqnarray}
which is of the same form as equation (\ref{eq1}) for the low-level excited-states.
Thus, the low-level populations satisfy equations of the same form as (\ref{eq6}).
This is a full solution for all levels, in the bundled-$n$ picture.
It includes direct couplings between the low-level populations (e.g. $n\rightarrow n^\prime$)
and indirect couplings via the high-level populations (e.g. $n\rightarrow\overline{n}\rightarrow n^\prime$).
However, we already have a description of the low-level problem in the fully-resolved picture,
given by equations (\ref{eq1}) and (\ref{eq6}). We can supplement these equations with the indirect couplings
of the bundled-$n$ picture, expanded over the low-level set using level weighting factors, $\omega_{ij}$. 
This projection corresponds to solving equations (\ref{eq1}) and (\ref{eq6}) with $C$, $S$, $R$ and $Q$ 
(including their appearance in equation (\ref{eq51})) replaced by ${\cal C}$, ${\cal S}$, 
${\cal R}$ and ${\cal Q}$ where
${\cal C}=C+\,^{{\rm ind}}\!C$, ${\cal S}=S+\,^{{\rm ind}}\!S$, ${\cal R}=R+\,^{{\rm ind}}\!R$ and
${\cal Q}=Q+\,^{{\rm ind}}\!Q$, where
\begin{eqnarray}
^{{\rm ind}}\!C^{z}_{\rho\sigma}&=&\omega_{\rho,\tau n}\omega_{\tau n^\prime,\sigma}
\sum_{\overline{n},\overline{n}^{\prime}}C^{z}_{\tau n,\tau\overline{n}^{\prime}}
\left(C^{z}\right)^{-1}_{\tau\overline{n}^{\prime},\tau\overline{n}}
C^{z}_{\tau\overline{n},\tau n^\prime}\, , \label{eq8}\\
^{{\rm ind}}\!S^{z}_{\nu^\prime i}&=&\omega_{\nu n,i}
\sum_{\overline{n},\overline{n}^{\prime}}S^{z}_{\nu^\prime,\nu\overline{n}^{\prime}}
\left(C^{z}\right)^{-1}_{\nu\overline{n}^{\prime},\nu\overline{n}}
C^{z}_{\nu\overline{n},\nu n}\, , \label{eq9}\\
^{{\rm ind}}\!R^{z+1}_{i\nu^\prime}&=&\omega_{i,\nu n}
\sum_{\overline{n},\overline{n}^{\prime}}C^{z}_{\nu n,\nu\overline{n}^{\prime}}
\left(C^{z}\right)^{-1}_{\nu\overline{n}^{\prime},\nu\overline{n}}
R^{z+1}_{\nu\overline{n},\nu^\prime}
\label{eq10}
\end{eqnarray}
and
\begin{eqnarray}
^{{\rm ind}}\!Q^{z}_{\rho\sigma}&=&\omega_{\rho,\tau n}\omega_{\tau n^\prime,\sigma}
\sum_{\overline{n},\overline{n}^{\prime}}S^{z-1}_{\rho,\tau\overline{n}^{\prime}}\left(C^{z-1}
\right)^{-1}_{\tau\overline{n}^{\prime},\tau\overline{n}}
R^{z}_{\tau\overline{n},\sigma}\, .
\label{eq11}
\end{eqnarray}
Although we now have a complete solution in terms of the fully-resolved
low-level and bundled-$n$ high-level picture, one further step is of practical
significance. In order to span a wide range of electron densities it is
necessary to treat very large principal quantum numbers in order to reach the
collision limit at low densities. It is not necessary to treat each $n$
individually. Rather, a set of representative  $n$-values can be used instead.
If $N_{\overline{n}}$ denotes the bundled-$n$ populations for $\overline{n}=n_{{\rm c}}+1,
n_{{\rm c}}+2, \ldots$, and ${\cal N}_{\overline{\overline{n}}}$ denotes a subset of them, then the
two are related via 
$N_{\overline{n}}=\omega_{\overline{n}\,\overline{\overline{n}}}{\cal N}_{\overline{\overline{n}}}$, where
$\omega_{\overline{n}\,\overline{\overline{n}}}$ are the interpolation coefficients. Substituting for
$N_{\overline{n}}$ into equations (\ref{eq1}) and (\ref{eq6}) yields a condensed set of equations for
${\cal N}_{\overline{\overline{n}}}$ and the ${\cal C}$, ${\cal S}$, ${\cal R}$ and ${\cal Q}$ obtained from
the condensed set of equations are identical in form to those obtained from the
full set of equations.

We note here that we have made an assumption viz. that, following dielectronic
capture, the autoionizing state is not perturbed by a further collision before it
either autoionizes or radiates. This is not due to a limitation of ADAS but rather a
choice  that we have made (and defined within the {\it adf09} specification) so as to
make the  general collisional--radiative problem tractable over a wide range of
electron densities.  Working explicitly with autoionization and radiative rates and
bound and non-bound states rather than partial dielectronic recombination rate 
coefficients and (mostly) bound states vastly increases the data requirements, in
general. Although our collisional--radiative model goes over to the correct LTE limit
at high electron densities, there is a density range ($\gtrsim 10^{16}$ cm$^{-3}$, found in 
laser-produced plasmas) where the levels of spectroscopic interest have non-LTE 
populations that are influenced by non-LTE populations of autoionizing levels  that
are themselves collisionally redistributed.  We describe an approximate solution in
section \ref{burgess-bethe} below.

We are now in a position to spell out our requirements of recombination data:
\newline
(i) We require recombination data from all metastable levels, not just the ground.
\newline
(ii) We require recombination data into particular final states.
\newline
(iii) We require the final-state to be level-resolved for $n\le n_{{\rm c}}$ and
parent-level-resolved bundled-$n$ for $n>n_{{\rm c}}$.
\newline
(iv) Parent metastable cross-coupling means that we require recombination into
metastable autoionizing final states.
\newline
(v) We need only produce data for a representative set of $\overline{n}$.
\newline
The ADAS {\it adf09} data specification incorporates all of these requirements.
Finally, we note that use of total zero-density ground-state recombination rate
coefficients is quite unsafe for the collisional--radiative modelling of dynamic finite-density
plasmas.

\subsection{Dielectronic recombination rate coefficient modelling}
\label{dr-coefft-modelling}
We have already noted that the partial recombination rate coefficient ($R^{z+1}_{i\nu}$)
includes contributions from three-body, radiative and dielectronic recombination.
In ADAS, three-body recombination rate coefficients are obtained from electron-impact 
ionization rate coefficients, via detailed balance. This also ensures that the correct
Saha--Boltzman limit is reached at high electron densities. Since three-body recombination 
is separate from dielectronic and radiative recombination, it is not necessary to consider 
if further. However, quantum mechanically, dielectronic and radiative recombination are 
indistinguishable
processes which interfere with each other. In practice (see Pindzola  et al. 1992\cite{pin92}), this
interference is a very small effect and can safely be neglected for our purposes. This is 
the independent processes approximation whereby dielectronic and radiative recombination can
be considered separately and is the approach taken by the database aspect of ADAS. Separate 
data files exist for dielectronic ({\it adf09}) and radiative recombination ({\it adf08})and
they can be updated independently. Our focus is dielectronic recombination. Details of the 
ADAS data status for radiative recombination can be found in Summers (2001\cite{sum01}).

In the isolated resonance approximation, the partial dielectronic recombination
rate coefficient $\alpha^{z+1}_{i\nu}$ from an initial metastable state $\nu$ into a 
resolved final state $i$ of an ion $X^{+z}$ is given by
\begin{eqnarray}
\alpha^{z+1}_{i\nu}&=&\left({4\pi a^2_0 I_{\rm H} \over k_{\rm B} T_{\rm e}}\right)^{3/2}
\sum_j{\omega_{j} 
\over 2\omega_{\nu}}\,{\rm e}^{-E_c/k_{\rm B} T_{\rm e}}\nonumber \\
 &\times&{ \sum_{l}A^{{\rm a}}_{j \rightarrow \nu, E_cl} \, A^{{\rm r}}_{j \rightarrow i}
\over \sum_{h} A^{{\rm r}}_{j \rightarrow h} + \sum_{m,l} A^{{\rm a}}_{j \rightarrow m, E_cl}}\, ,
\label{eq12}
\end{eqnarray}
where $\omega_j$ is the statistical weight of the
$(N+1)$-electron doubly-excited resonance state $j$, $\omega_\nu$ is the statistical weight
of the $N$-electron target state and the autoionization ($A^{\rm a}$) and radiative
($A^{\rm r}$) rates are in inverse seconds. Here, $E_c$ is the energy of the continuum electron
(with angular momentum $l$), which is fixed by the position of the resonances, and
$I_{\rm H}$ is the ionization potential energy of the hydrogen atom (both in the
same units of energy), $k_{\rm B}$ is the Boltzman constant, $T_{\rm e}$ the electron temperature
and $(4\pi a^2_0)^{3/2}=6.6011\times10^{-24}$ cm$^3$. The effect of interacting resonances on 
dielectronic recombination has been investigated by Pindzola  et al. (1992\cite{pin92}) and can safely be
neglected, at least in the absence of external electric and magnetic fields (see section 3 below).
While autoionization rates can be determined (within the isolated resonance approximation)
via the fitting of resonances calculated in a close-coupling approximation, or via the
extrapolation of threshold close-coupling collision strengths using the correspondence principle,
it is usual now to introduce a further approximation -- that of using distorted waves, i.e.
the autoionization rates are calculated via perturbation theory using the Golden Rule
(Dirac 1930). This is the only approximation that we have made so far that may need
to be reconsidered in certain cases. In low-charge ions, a perturbative distorted wave
calculation may give inaccurate autoionization rates compared to those calculated in a 
close-coupling approximation. However, this only has a direct effect on the partial
dielectronic recombination rate if the autoionization rates do not `cancel-out' between
the numerator and denominator of (\ref{eq12}) -- typically, autoionization rates are orders of
magnitude larger than radiative rates.

One could obtain (some) partial dielectronic recombination data from an $R$-matrix
photoionization calculation, on making use of detailed balance, either in the
absence of radiation damping (Nahar \& Pradhan 1994\cite{nah94}) or with its inclusion
(Robicheaux  et al. 1995\cite{rob95}, Zhang  et al. 1999\cite{zha99}).
(One must take care not to double count the 
radiative recombination contribution in the modelling now.) However, this cannot provide
us with a complete set of partial recombination rate coefficients since it is
only possible to compute photoionization from (i.e. photorecombination to) a relatively
low-lying set of states -- up to $n\simeq 10$, say. Total recombination rates are obtained by
supplementing the photorecombination data with high-$n$ `close-coupling' dielectronic recombination
rate coefficients calculated using Bell \& Seaton (1985\cite{bel85}) or Hickman's (1984\cite{hic84})
approach. This is based on the radiative-loss term from a unitary S-matrix and
does not, and cannot, resolve recombination into a particular final state, which is
essential for collisional--radiative modelling. Indeed, even when summed-over all final
states, errors can still result. This has been demonstrated
explicitly by Gorczyca  et al. (2002\cite{gor02}) in the case of Fe$^{17+}$. They found that only
the IPIRDW approach could reproduce the measured dielectronic recombination cross section of Savin  et al.
(1997\cite{sav97}, 1999\cite{sav99}) for high Rydberg states. Thus, our initial goal is to
generate complete data sets within the independent processes and isolated resonance
using distorted waves (IPIRDW) approximation. Subsequently, selective 
upgrades from $R$-matrix data may be made via, for example, the RmaX network which
can be viewed as a progression of the Iron Project (Hummer  et al. 1993\cite{hum93}) and which is
focusing on X-ray transitions -- see, for example, Ballance  et al. (2001\cite{bal01}).
We note that while the Opacity Project (Seaton 1987\cite{sea87})
calculated a large amount of photoionization data, which in principle could be used for
recombination (via detailed balance), unfortunately, only total photoionization
cross sections were archived, i.e. summed-over the final electron continuum, and
so it is impossible to apply detailed balance and so it cannot be used as a source
of recombination data.

We use the code {\sc autostructure} (Badnell 1986\cite{bad86}, Badnell \& Pindzola 1989\cite{bad89},
Badnell 1997\cite{bad97}) to calculate multi-configuration intermediate coupling energy 
levels and rates within the IPIRDW approximation. The code can make use both
of non-relativistic and semi-relativistic wavefunctions (Pindzola and Badnell 1990\cite{pin90}).
The low-$n$ problem is no different from the one of
computing atomic structure. The high-$n$ problem requires some discussion. 
The mean radius of a Rydberg orbital scales as $n^2$
and so it rapidly becomes impossible to calculate an explicit bound orbital
(for $n>20$, say) and some approximation must be made. We note that (Seaton 1983\cite{sea83})
\begin{eqnarray}
\lim_{n \rightarrow \infty}\left({\pi n^3\over 2z^2}
\right)^{3/2}P_{nl}(r)=\left.F_{kl}(r)\right|_{k=0}\, ,
\label{eq13}
\end{eqnarray}
where the bound orbitals are normalized to unity and the continuum
orbitals to $\pi\delta(k-k')$, here $k^2=E_c$(Ry). The approach taken in
{\sc autostructure} is to make use of (\ref{eq13}) at finite $n$, i.e. to approximate
the bound orbital by a suitably normalized zero-energy continuum orbital,
for $n>15+l^2/4$ (evaluated in integer arithmetic). A further refinement
is to evaluate the (true) continuum orbital for the incident electron at
$E_c+z^2/n^2$ Rydbergs (instead of $E_c$) so as to maintain the same
transition energy.
(In the Bethe approximation, the free--free dipole acceleration integral is 
proportional to $\Delta\varepsilon^2$ times a slowly-varying-with-energy 
dipole-length integral, where $\Delta\varepsilon$ is the transition energy.)
Actually, the true continuum orbital is calculated at about 15 energies
per $l$ and the one- and two-body bound--free integrals are interpolated at the required
energy, which is given by energy conservation. (The use of a zero-energy continuum
orbital means that long-range free--free integrals arise and these are
treated using the techniques of distorted wave scattering theory, see Badnell 1983.)
For each $nl$, {\sc autostructure} reforms both
the $N$- and $(N+1)$-electron Hamiltonians and diagonalizes them (separately)
to reform the rates. Only for H-like ions have we found it
necessary to treat all $l$ at the same time, and treat only each $n$ separately.
Typically, each $n$ is calculated explicitly until no new continua can 
open-up and then only a representative set of $n$, up to $n=999$, is used.
This approach avoids the extrapolation of low-$n$ autoionization rates or, even worse,
partial dielectronic recombination rate coefficients to high-$n$. Use of (\ref{eq13})
is, in effect, an interpolation since it is exact in the limit $n\rightarrow\infty$
and any error is bounded at the lowest-$n$ by knowledge of the `exact' result
obtained from using $P_{nl}$ directly, rather than $F_{kl}$. 

{\sc autostructure} is implemented within ADAS as ADAS701. It produces the
raw autoionization and radiative rates. To produce partial dielectronic
recombination rate coefficients, according to the prescription of Sect. 2.1,
requires further non-trivial organization of the raw data. In particular, 
radiative transitions between
highly-excited Rydberg states are computed hydrogenically and added-in
during a `post-processing' exercise with the code {\sc adasdr}, which is
implemented with ADAS as ADAS702. Also, observed energies for the core 
and parent levels are used at this stage to ensure accurate positioning of the
resonances and, hence, accurate low-temperature rate coefficients. {\sc adasdr}
outputs directly the {\it adf09} file for use by ADAS. Separate {\it adf09} files
are produced for different `core-excitations' ($n\rightarrow n'$), e.g. $1\rightarrow2$,
$2\rightarrow2$ and $2\rightarrow3$ for Li-like ions. This enables selective upgrades
of the {\it adf09} database.

\subsection{Exploitation of the Burgess--Bethe approach}
\label{burgess-bethe}

We are concerned with how the precise calculations described above relate to other calculations and, 
in particular, to those commonly used in astrophysics. Our {\it baseline}  calculation is
based on the methodologies of Burgess, which represent what can be achieved without
recourse to the detail of the above sections. The Burgess  GF itself was in fact a functional fit to
extended numerical calculations.  The associated code, with extensions, we call the
`Burgess--Bethe general program' (BBGP).  It will be shown in this sub-section how the BBGP can
be used to obtain a working model for the $l$-redistribution of doubly-excited states and, hence, 
provide a correction to accurate, but unredistributed, dielectronic data so as to model the dynamic
part of the plasma microfield.  Also, our {\it baseline}, calculated using the BBGP, will allow an 
assessment of the typical error that is present in the general dielectronic modelling in astrophysics 
to date. 

In the LS-coupled term picture, introduce a set $P$ of parent terms $\gamma_p~^{2S\!_p+1}L_p$ of
energy $E_p$ relative to the ground parent term, indexed by $p$.  Suppose that the excited parents
are those with $p_{min} \leq p \leq P$.  The metastable parents, which are the initial metastables
for recombination and the final parents on which the recombining excited $nl$-electron is built,
are the subset $1 \leq p < p_{min}=M$.  Let $p'$ denote an initial parent with the incident
electron denoted by $k'l'$.  We wish to re-establish the expressions used by Burgess in his
development and it is helpful to work in $z$-scaled dimensionless coordinates.  Then, introducing
$z_{\rm eff}$, the effective charge, the collision strength for a dipole excitation of
$\gamma_p~^{2S\!_p+1}L_p$  to $\gamma_p'~^{2S\!_p+1}L_p'$, evaluated in the Bethe approximation, is
given by
\begin{eqnarray}
&&\Omega((S\!_pL_p')k'l',(S\!_pL_p)kl SL) = 48((z_{\rm eff}+1)^2/z_{\rm eff}^4) \nonumber \\
&&\times\left(\frac{I_{\rm H}}{\Delta \epsilon_{pp'}}\right) \left(\frac{(2S+1)(2L+1)}{2(2S\!_p+1)}\right)l_>
\left\{ \begin{array}{ccc}
L_p' & L_p & 1 \\
l  & l'    & L \end{array} \right\}^2 \nonumber \\
&&\times(2S\!_p+1)(2L_p'+1)f_{S\!_pL_p' \rightarrow S\!_pL_p}
\left|\left<F_{k'l'}\left|\rho^{-2}\right|F_{kl}\right>\right|^2      
\label{eq14}
\end{eqnarray}
\noindent in terms of the parent oscillator strength $f$, where $l_>={\rm max}(l,l')$.  
For dielectronic recombination, it is
convenient to express this in terms of the Einstein A-coefficient for the parent
transition and to analytically continue the collision strength to negative energies for the
$kl$ electron, that is as $\kappa=k/z_{\rm eff} \rightarrow i/n$.  Formally, we identify a
band of free-electron energies ${\rm d} E/I_{\rm H}=z_{\rm eff}^2{\rm d}\epsilon/I_{\rm H}$, with the separation
between $n$-shells $2z_{\rm eff}^2I_{\rm H}/n^3$, and let
\begin{eqnarray}
\left|\left<F_{k'l'}\left|\rho^{-2}\right|F_{kl}\right>\right|^2{\rm d}(\epsilon/I_{\rm H})    
\rightarrow (\pi/16)(\epsilon/I_{\rm H})^4\left|\left<F_{k'l'}\left|\rho\right|P_{nl}\right>\right|^2\, .
\label{eq15}
\end{eqnarray}
\noindent Then, the (dielectronic) resonance-capture cross section is given by        
\begin{eqnarray}
&&Q^{\rm c}((S\!_pL_p')k'l',(S\!_pL_p)kl SL){\rm d}(E/I_{\rm H}) \rightarrow
\left(\frac{(z_{\rm eff}+1)^2}{z_{\rm eff}^4}\right)  \nonumber \\
&&\times \left(\frac{6\pi^2 a_0^3}{\alpha^4c}\right ) \left(\frac{\Delta \epsilon_{pp'}}{I_{\rm H}}\right)
\left(\frac{I_{\rm H}}{\epsilon}\right)\left(\frac{(2S+1)(2L+1)}{2(2S\!_p+1)}\right) l_>  \nonumber \\
&&\times\left\{ \begin{array}{ccc}
L_p' & L_p & 1 \\
l  & l'    & L \end{array} \right\}^2 \left(\frac{A^{\rm r}(S\!_pL_p \rightarrow S\!_pL_p')}
{(z_{\rm eff}+1)^4}\right ) 
\left|\left<F_{k'l'}\left|\rho\right|P_{nl}\right>\right|^2    \, . 
\label{eq16}
\end{eqnarray}

\noindent The inverse (Auger) rate coefficient is obtained by
invoking detailed balance as     
\begin{eqnarray}
&&A^{\rm a}((S\!_pL_p)nl SL \rightarrow (S\!_pL_p')k'l') = \left(\frac{(z_{\rm eff}+1)^2}{z_{\rm eff}^2}\right)
\left(\frac{3}{2\alpha^3}\right) \nonumber \\
&&\times\left(\frac{\Delta \epsilon_{pp'}}{I_{\rm H}}\right)
 (2L_p'+1)l_> \left\{ \begin{array}{ccc}
L_p' & L_p & 1 \\
l  & l'    & L \end{array} \right\}^2 \left(\frac{A^{\rm r}(S\!_pL_p \rightarrow S\!_pL_p')}
{(z_{\rm eff}+1)^4}\right)   \nonumber \\
&&\times\left|\left<F_{k'l'}\left|\rho\right|P_{nl}\right>\right|^2  \, .
\label{eq17}
\end{eqnarray}

\noindent For the generalized collisional--radiative modelling of light element ions, it is convenient to use
$L$-averaged doubly-excited levels, but still resolved by spin $S$, whereas the BBGP treatment uses
$SL$-averaged levels.  The corresponding Auger rates are then 
\begin{eqnarray}
&&A^{\rm a}((S\!_pL_p)nl S \rightarrow (S\!_pL_p')k'l') = \left(\frac{(z_{\rm eff}+1)^2}
{z_{\rm eff}^2}\right)\left(\frac{1}{2\alpha^3}\right) 
\left(\frac{\Delta \epsilon_{pp'}}{I_{\rm H}}\right)
\nonumber \\
&&\times
\left(\frac{l_>}{(2l+1)}\right) \left(\frac{A^{\rm r}(S\!_pL_p \rightarrow S\!_pL_p')}
{(z_{\rm eff}+1)^4}\right) 
%  \nonumber \\
\left|\left<F_{k'l'}\left|\rho\right|P_{nl}\right>\right|^2  \, ,
\label{eq18}
\end{eqnarray}
\noindent in both cases.  The matching resonance-capture coefficients are obtained by detailed balance,
or by summing and averaging over the resolved expression for $Q^{\rm c}$ given above.

Turning to the radiative decay of the doubly-excited resonant states in the $LS$-resolved picture: the
spontaneous emission coefficient, with a passive spectator  in the $nl$ shell, is given by
\begin{eqnarray}
&&A^{\rm r}((S\!_pL_p)nl SL \rightarrow (S\!_pL_p')nlSL') = (2L_p+1)(2L'+1) \nonumber \\
&&\times\left\{ \begin{array}{ccc}
L_p' & L_p & 1 \\
L  & L'    & l \end{array} \right\}^2 
A^{\rm r}(S\!_pL_p \rightarrow S\!_pL_p')  \, ,
\label{eq19}
\end{eqnarray}
\noindent where $A^{\rm r}(S\!_pL_p \rightarrow S\!_pL_p')$ is the parent-core spontaneous transition probability.
The $L$- and $LS$-averaged probabilities  are both simply equal to $A^{\rm r}(S\!_pL_p \rightarrow S\!_pL_p')$.  
The BBGP method exploits the fact that, in the dipole case, 
\begin{eqnarray}
\frac{A^{\rm a}}{A^{\rm r}}&=&\frac{l_>\left|\left<F_{k'l'}\left|\rho\right|P_{nl}\right>\right|^2}
{2\alpha^3(2l+1)(z_{\rm eff}+1)^2z_{\rm eff}^2}    \, ,
\label{eq20}
\end{eqnarray}
\noindent and efficient recurrence relations are available for the generation of hydrogenic bound--free radial
integrals for all parameter values.  It is clear that the Bethe approximation for the partial collision
strengths can be substantially in error for $0 \leq l \lesssim 2$.  Burgess introduced correction factors for
the lowest partial collision strengths, based-on a comparison with more sophisticated collision calculation
results that were available at the time.  More precisely, introduce 
\begin{eqnarray}
{\it cor}_l & = & \left.\sum_{l'}\Omega((S\!_pL_p')k'l',(S\!_pL_p)k|_{=0}l)\right/ \nonumber \\
&&\sum_{l'}\Omega^{\rm Bethe}((S\!_pL_p')k'l',(S\!_pL_p)k|_{=0}l) \, .
\label{eq21}
\end{eqnarray}

The general formula for zero-density total dielectronic recombination rate coefficients 
used a fixed-set of ${\it cor}_l$ for all parent transitions.
These were based-on parent $1{\rm s} \rightarrow 2{\rm p}$ transitions and it is the case that
the inclusion of corrections is most significant for parent $\Delta n \geq 1$ transitions.  
To exploit the BBGP method
beyond its use for the general formula, we must establish the population equations of the doubly-excited
levels.  For $LS$-averaged levels, the number densities expressed in terms of their deviations,
$b_{p,nl}$, from Saha--Boltzmann, and referred to the initial parent $p'$, are given by
\begin{eqnarray}
N_{p,nl} = N_{\rm e} N^{+}_{p'}8\left[\frac{\pi a_0^2I_{\rm H}}{k_{\rm B} T_{\rm e}}\right]^{3/2}
\frac{\omega_{p,nl}}{\omega_p'}e^{-E/kT_{\rm e}}b_{p,nl}\, .
\label{eq22}
\end{eqnarray}

\noindent Then, in the BBGP {\it baseline} zero-density limit, with only resonant capture from the
$p'$ parent balanced by Auger breakup and radiative stabilization back to the same parent, we have
\begin{eqnarray}
b_{p,nl} & = & \left(\frac{\sum_{l'}A^{\rm a}(p,nl \rightarrow p'k'l')}{\sum_{l'}A^{\rm a}(p,nl
\rightarrow p'k'l')+A^{\rm r}(p,nl \rightarrow p',nl)}\right) \, .
\label{eq23}
\end{eqnarray}
In the extended BBGP program, we can also include resonant-capture from initial metastables other
than the ground, dipole-allowed collisional redistribution between adjacent doubly-excited
$l$-substates by secondary ion- and electron-impact, and losses by `alternate' Auger break-up
and parent radiative transition pathways.  The population equations for the $l$-substates of
a doubly-excited $n$-shell become   
\begin{eqnarray}
&&-\left(N_{\rm e} q^{\rm e}_{nl-1 \rightarrow nl}+N^{z_{\rm eff}}q^{z_{\rm eff}}_{nl-1 \rightarrow nl}\right)
N_{p,nl-1}\nonumber \\
&&+\left(\sum_{l'=l\pm1}N_{\rm e} q^{\rm e}_{nl \rightarrow nl'}
+\sum_{l'=l\pm1}N^{z_{\rm eff}}q^{z_{\rm eff}}_{nl \rightarrow nl'} \right.\nonumber \\
&&+\left.\sum_{p_1=1}^{p-1}\sum_{l'=l-1}^{l+1}A^{\rm a}_{p,nl \rightarrow p_1,\kappa
 l'}+\sum_{p_1=1}^{p-1}A^{\rm r}_{p,nl \rightarrow p_1,nl}\right)N_{p,nl}\nonumber \\
&& -\left(N_{\rm e} q^{\rm e}_{nl+1 \rightarrow nl}+N_{z_{\rm eff}}q^{z_{\rm eff}}_{nl+1 \rightarrow nl}\right)
N_{p,nl+1} \nonumber \\
&&= N_{\rm e}\sum_{p_2=1}^{M}\sum_{l'=l-1}^{l+1}q^{\rm c}_{p_2,\kappa l' \rightarrow p,nl}N_{p_2} 
+\sum_{p_1=p+1}^{P}A^{\rm r}_{p_1,nl \rightarrow p,nl}N_{p_1,nl}\, .
\label{eq24}
\end{eqnarray}
\noindent These equations may be solved progressively downwards through the levels built on
excited parents, terminating with levels built on the `final' ground and metastable parents.
The calculations yield state-selective dielectronic recombination coefficients to levels built on each
metastable parent, together with Auger rates, from levels built on metastables above the 
ground parent, to lower metastable parents.  The results from the above solution, at zero density, we
call {\it baseline} data.            

\noindent We note some details of the implementation:

\subsubsection{Collisional rates}
Ion and electron $l$-redistributive cross sections are evaluated following the method of
Pengelly \& Seaton (1964\cite{pen64}).  This is to be viewed as a very simplified treatment of the
dynamic part of the plasma microfield associated with ions which move closer than
their neighbours to the target.  The quasi-static part of the ion microfield, which is
approximately 1/3 of the total, is ignored at this level of analysis (see section
\ref{experimental-validation} for further discussion of field effects in an experimental
context).     
\subsubsection{Energy levels}
The energy differences  $E_{p,nl~S}-E_{p,nl\pm 1~S}$ are critical to the problem of  doubly-excited-state
redistribution. They are small, tending to zero at large $l$.  Thus, the redistributive cross-sections are
very large, remaining finite in the degenerate level limit only because of the finite radiative lifetime of
the target (which is short for resonant states) or through screening of the projectile by nearest
neighbours.   For the $l=0,1,2$ waves of the $n$-shell spectator electron, quantum defect expansions of the
form $\mu_{p,nl~S}=a_0+a_1/n^2$ are usually available.  For the higher $l (>2)$, the large number of
energies required can be estimated more economically from the dipole polarizabilities, $\alpha^{{\rm
pol}}_p$, of the parents, following Edlen(1964), as
\begin{eqnarray}
E_{p,nl} & = &(z_1/n)^2[1+5.23504^{-4}(z_1/n)^2(n/(l+1/2)+ \nonumber \\
& & 3/4)]+\alpha^{\rm pol}_p(z_1/n)^4(3n^2-l(l+1))/ \nonumber \\
& & (2n(l-1/2)l(l+1/2)(l+1)(l+3/2))     \, .
\label{eq25}  
\end{eqnarray}
Small-scale {\sc autostructure} runs have been used to prepare these data for the {\it baseline}.  
We use two-term quantum defect expansions fitted at $n \rightarrow \infty$ and $n = 10$.
\begin{figure}[!htp]
\centering
\ \psfig{file=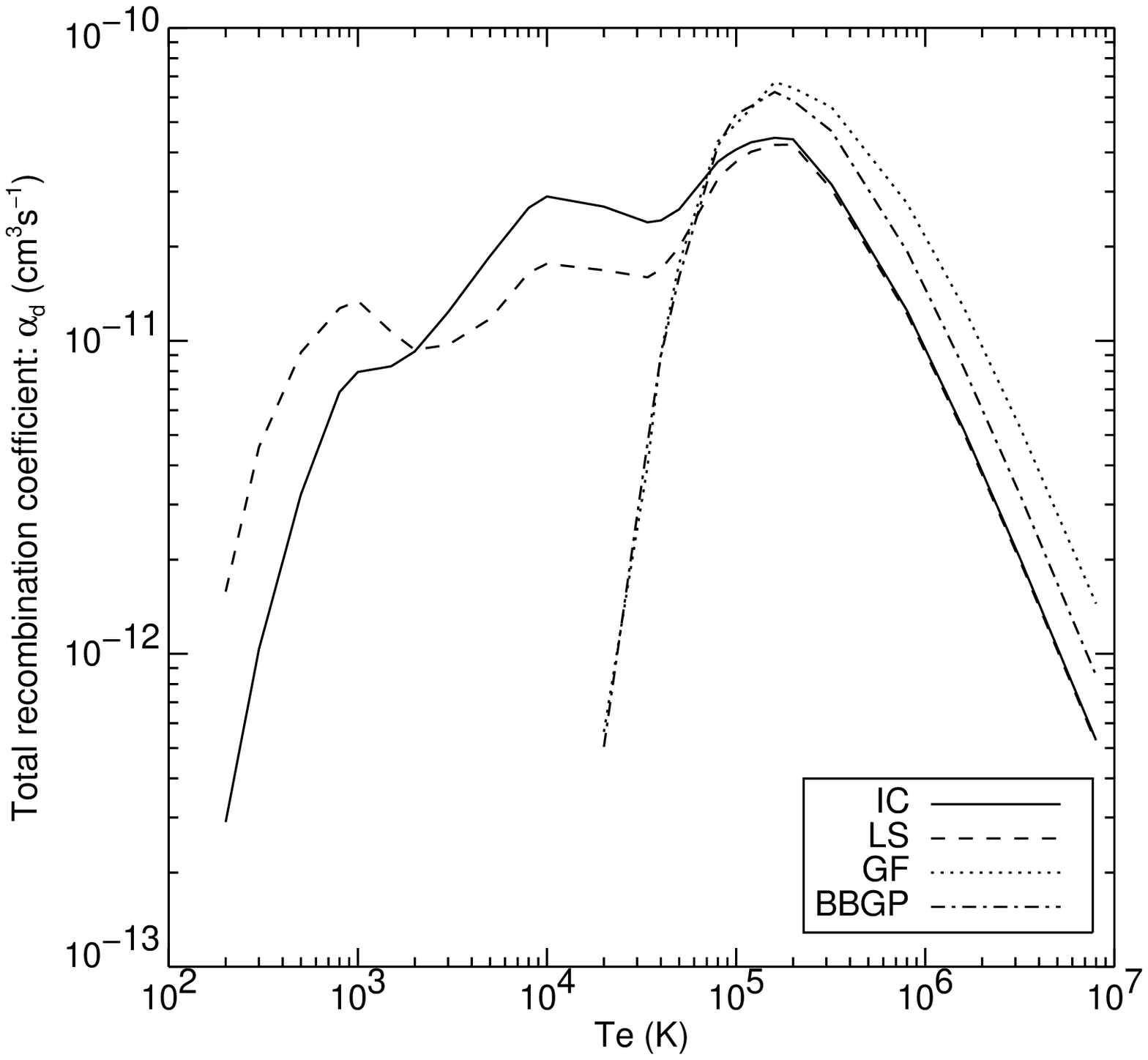,width=8.5cm}
\ \psfig{file=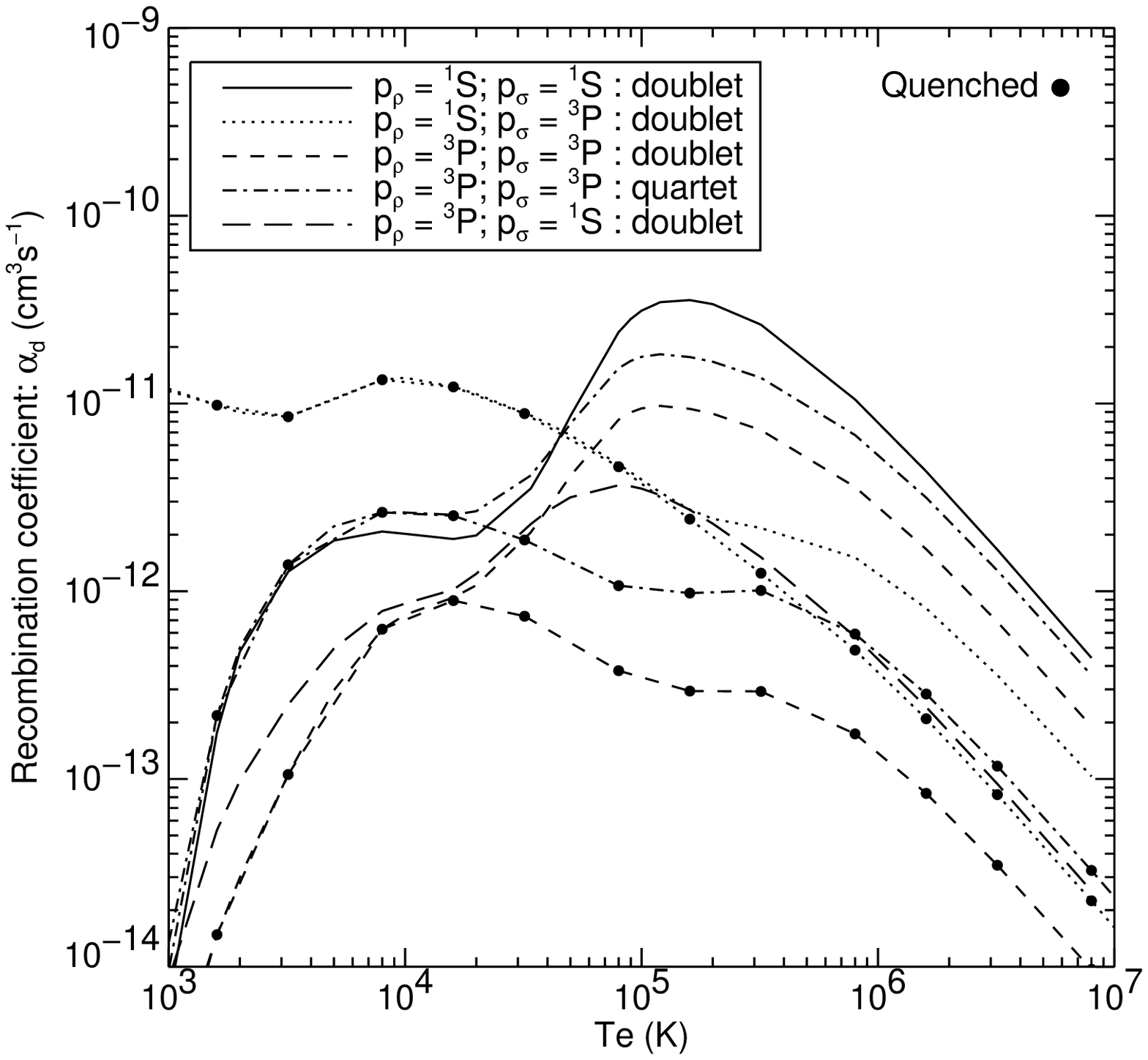,width=8.5cm}
 \caption{The graphs contrast $\alpha_{\rm d}^{({\rm tot})}(p_{\rho} \rightarrow p_i
\rightarrow p_{\sigma})$ vs $T_{\rm e}$  for the dielectronic recombination of O$^{4+}$ ions in various
approximations at zero density. (a) $p_{\rho}$= O$^{4+}(2{\rm s}^2~^1{\rm S})$, $p_{\sigma}$=
O$^{3+}(2{\rm s}^22{\rm p}~^2{\rm P})$: $\Delta n = 0$ and $\Delta n = 1$ intermediate parents included.  The 
Burgess GF includes only the dipole $2{\rm s}^2~^1{\rm S}-2{\rm s}2{\rm p}~^1{\rm P}$ and
$2{\rm s}^2~^1{\rm S}-2{\rm s}3{\rm p}~^1{\rm P}$ transitions.  BBGP incorporates specific ${\it cor}_l$ correction factors,
energies which differ from those implicit in the GF, and the $3{\rm p}-3{\rm s}$ alternative Auger channel.  
The {\it level 1}  and {\it level 2}  results include all allowed and non-allowed parent
transitions within the $n=2$ and $n=3$ complexes. Note the low temperature extension which cannot be
modelled with the GF and BBGP.  The correct distinction and positioning of the key lowest resonances are
possible only at {\it level 2}. (b) $p_{\rho}$= O$^{4+}(2{\rm s}^2~^1{\rm S})$ and O$^{4+}(2{\rm s}2{\rm p}~^3{\rm P})$,
$p_{\sigma}$=O$^{3+}(2{\rm s}^22{\rm p}~^2{\rm P})$ and O$^{3+}(2{\rm s}2{\rm p}^2~^4{\rm P})$: {\it level 1} results
separated by spin-system and final parent.  Recombination from-and-to metastables cannot be handled by the GF. 
Excited-states built on the $2{\rm s}2{\rm p}~^3{\rm P}$ parent have a spin-change autoionization
pathway.  The {\it level 1} metastable-resolved totals do not include this loss.  Within an LS-coupled GCR
picture using {\it level 1} data, spin-breakdown Auger data is included explicitly in the GCR calculations
for the correct linking of systems built on the $2{\rm s}^2~^1{\rm S}$ and $2{\rm s}2{\rm p}~^3{\rm P}$ parents. The
relevant final-parent-changing Auger data is included in the {\it adf09} data file specification.  For
comparison with simpler treatments, totals including quenching of $n$-shells $>$ 4  built on the
$2{\rm s}2{\rm p}~^3{\rm P}$ parent are also shown.}
 \label{fig:fig1-new} 
\end{figure}
\begin{figure}[!htp]
\centering
\ \psfig{file=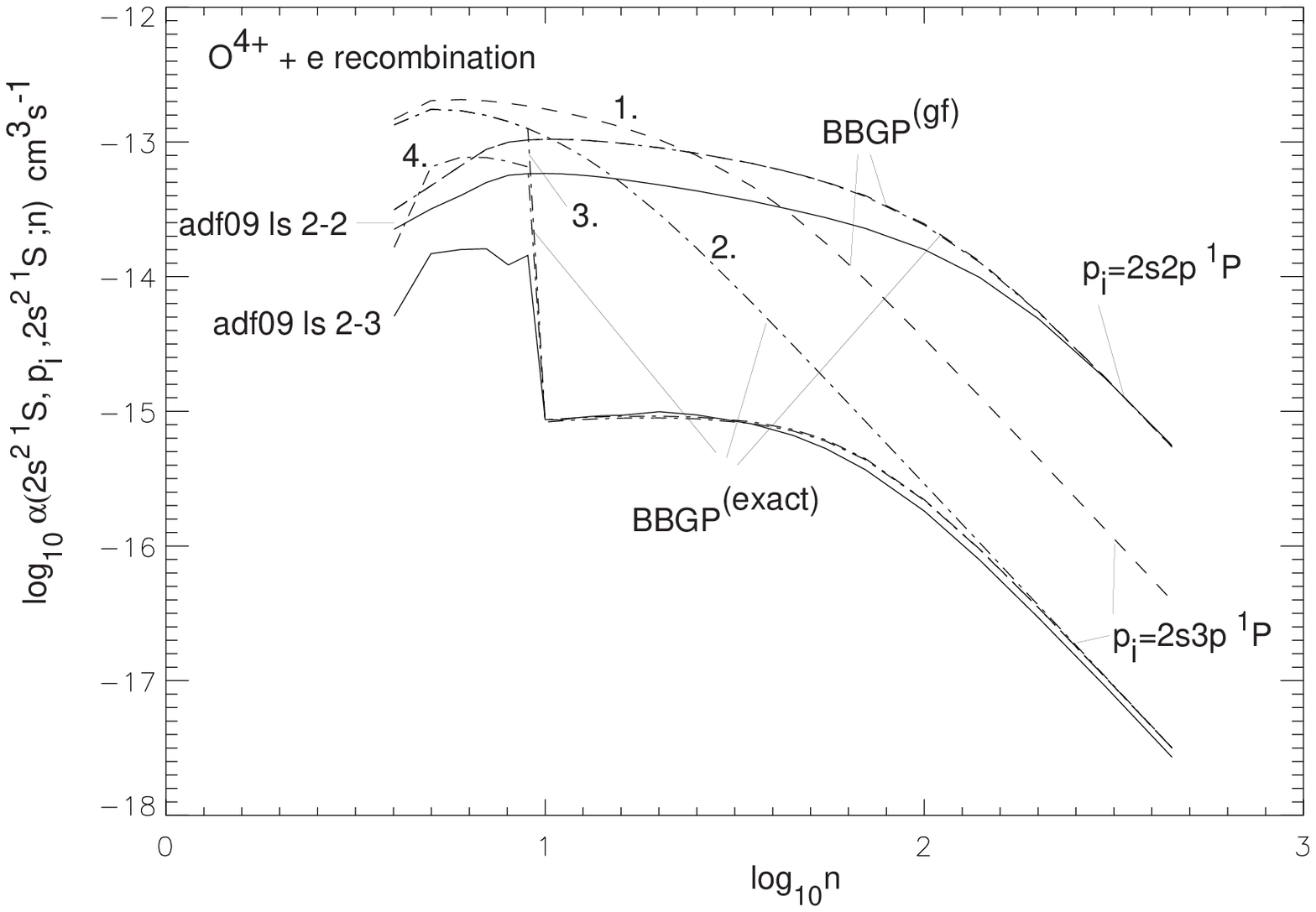,width=8.5cm}
\ \psfig{file=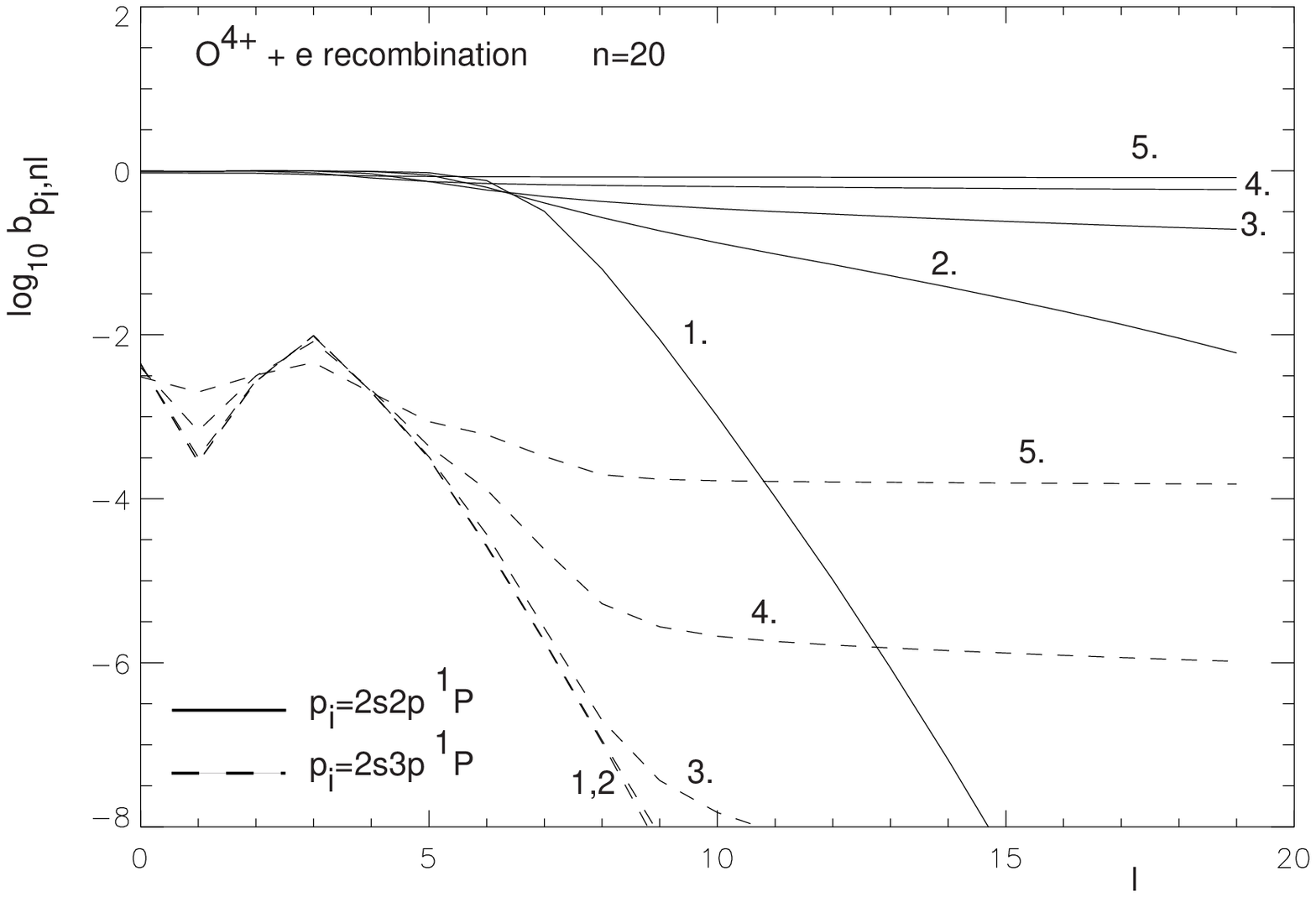,width=8.5cm}
 \caption{O$^{4+}$ recombination. (a) Partial $n$-shell recombination coefficients:
initial recombining metastable $2{\rm s}^2~^1{\rm S}$, intermediate excited parent $p_i$, final states
$(2{\rm s}^2~^1{\rm S})n$ with $p_i= 2{\rm s}2{\rm p}~^1{\rm P}$ and $2{\rm s}3{\rm p}~^1{\rm P}$.  
$T_{\rm e}=\rm{1.6}\times \rm{10}^6$K,
$N_{\rm e}=N_{\rm p}=\rm0$.  BBGP$^{\rm (gf)}$ indicates use of standard $cor_l$ matching the GF, BBGP$^{\rm(exact)}$ indicates
use of specific $cor_l$. For the $\Delta n = 0$ case, BBGP$^{\rm(gf)}$ and BBGP$^{\rm(exact)}$ are superposed. 
$\Delta n = 1$ cases: 1. no alternative Auger channels, GF $cor_l$;  2. no alternative Auger channels,
specific $cor_l$; 3. $2{\rm s}3{\rm p}~^1{\rm P}$ -- $2{\rm s}3{\rm s}~^1{\rm S}$ alternative Auger channel, specific $cor_l$;
4. as before and $2{\rm s}3{\rm p}~^1{\rm P}$ -- $2{\rm s}2{\rm p}~^1{\rm P}$ alternative Auger channel, specific $cor_l$.  (b)
$b_{p_i,nl}$ factors for doubly-excited states of O$^{3+}$ relative to O$^{4+}$ $2{\rm s}^2~^1{\rm S}$ for $p_i=
2{\rm s}2{\rm p}~^1{\rm P}$ and $2{\rm s}3{\rm p}~^1{\rm P}$, $n=20$, 
$T_{\rm e}=\rm{10}^6$K, $N_{\rm e}=N_{\rm p}$, $Z_{\rm eff}=1$ . Cases: 1.
$N_{\rm e}=\rm{10}^{10}$ cm$^{-3}$; 2. $N_{\rm e}=\rm{10}^{12}$ cm$^{-3}$; 3. $N_{\rm e}=\rm{10}^{13}$ cm$^{-3}$; 4.
$N_{\rm e}=\rm{10}^{14}$ cm$^{-3}$; 5. $N_{\rm e}=\rm{10}^{15}$ cm$^{-3}$. Note the alternative Auger channel
reduction for the $p_i= 2{\rm s}3{\rm p}~^1{\rm P}$ graphs.}
 \label{fig:fig2-new} 
\end{figure}

\subsubsection{Bethe correction factors and radiative transition probabilities}
For our comparisons, it is important that the inputs for the BBGP {\it baseline}
calculations are consistent with the {\sc autostructure} 
calculations for the {\it level 1} and {\it level 2} results. Small-scale
{\sc autostructure} runs have been used to prepare these inputs for the {\it baseline}. 
Low partial-wave collision strengths are now widely available (e.g. from {\sc autostructure}), 
such that the ${\it cor}_l$ can be prepared fairly easily. Within the ADAS Project, efficient
subroutines for bound--free integrals and complete BBGP
calculations are available together with sets of ${\it cor}_l$  for
the principal types of transitions.  They are available to those for whom direct
utilization of {\it level 1} and {\it level 2} data in full GCR modelling is not an
option.    

The results presented in this overview paper are illustrative only. Fig.~\ref{fig:fig1-new} contrasts 
zero-density total dielectronic recombination coefficients ($\alpha^{\rm tot}_{\rm d}$) calculated in the GF
and the BBGP {\it baseline} approximations with those of the {\it level 1} and {\it level 2} computations 
reported here.  In particular, we note that {\it level 1} and 2 data are required to describe the recombination at
low temperatures and that the {\it level 2} data provides a noticeable refinement over the {\it level 1} results.
Figs.~\ref{fig:fig2-new}a,b illustrate the partial recombination into $n$-shells and the population
structure of the $l$-subshells of a representative doubly-excited $n$-shell.  Fig.~\ref{fig:fig2-new}a
shows the very good convergence of BBGP to {\it level 1} data with increasing completeness of
alternate Auger pathways.  Fig.~\ref{fig:fig2-new}b shows the effects of collisional redistribution at
finite-density.  A ratio of the sum over $l$-substates at a given density to that at zero density yields a
BBGP finite-density adjustment factor of the total $n$-shell capture at zero density.  The consistency
between the BBGP, {\it level 1}, and {\it level 2} approaches allows us to use this adjustment factor on the {\it level 1} 
and {\it level 2} data. In advanced generalized collisional--radiative modelling, the BBGP finite-density
redistributive code acts as an interface between the extraction of state-selective zero-density
dielectronic data from the ADAS {\it adf09} database and its entry into the GCR population codes, corrected
for finite-density doubly-excited state redistribution. Note also that routine semi-automatic comparisons
as, illustrated here, provide the theoretical uncertainty estimate with which we can tag each
dielectronic datum.  

\begin{figure}[!htp]
\centering
\ \psfig{file=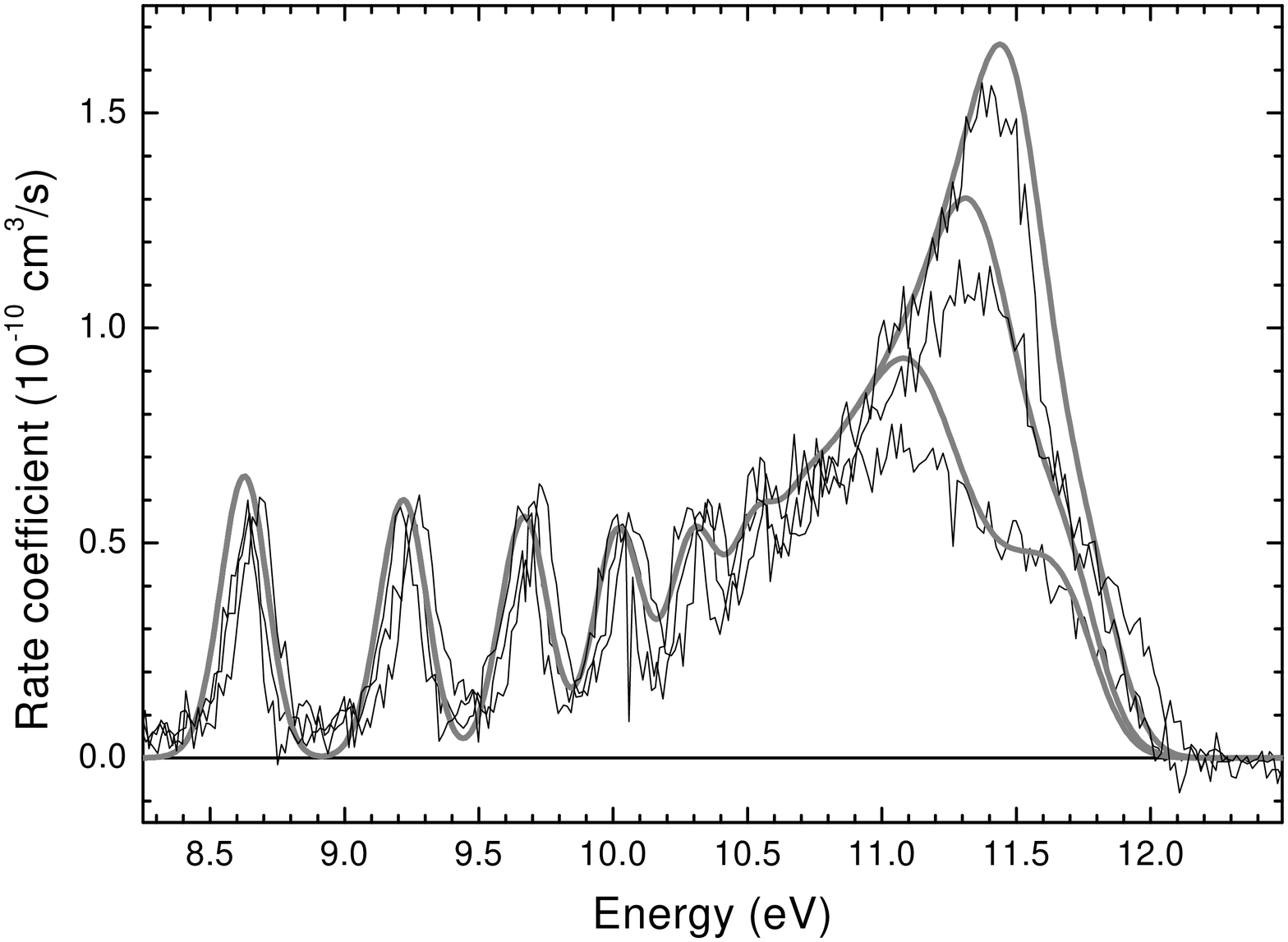,width=8.5cm}
\ \psfig{file=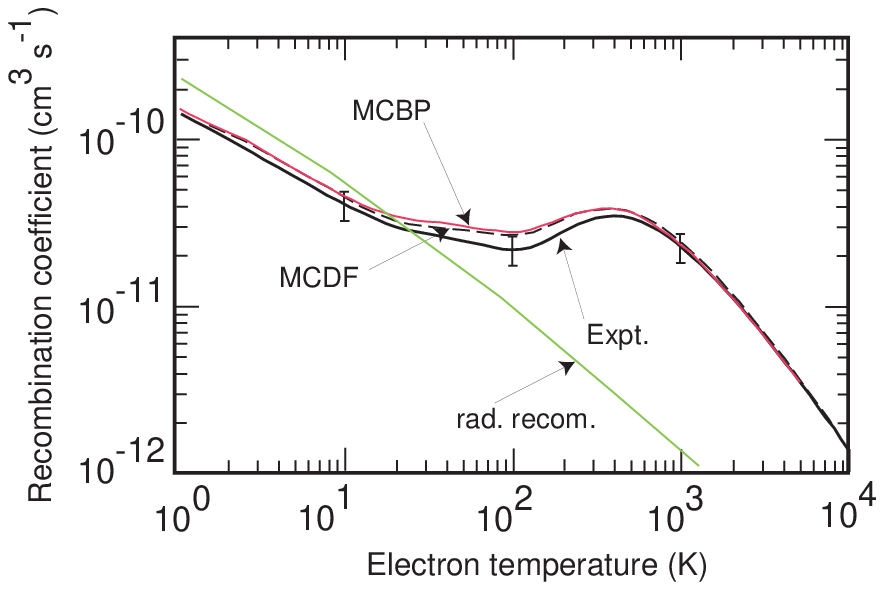,width=8.5cm}
 \caption{(a) A theoretical reconstruction is shown of the observed 
dielectronic recombination resonances for the light ion O$^{5+}+{\rm e}^- \rightarrow $O$^{4+}$.
The measurement is from the CRYRING heavy-ion storage ring (B\"{o}hm  et al. 2002\cite{boh02}).
Note that the dielectronic recombination data from the methods described in section 
\ref{dr-coefft-modelling} have been convoluted with the experimental velocity distribution,
which has the units of a rate coefficient.  The three pairs of curves correspond to three
different beam energies which results-in three different cut-offs in the maximum principal
quantum number detected. The experimental and theoretical data agree quite nicely.
(b) The Maxwellian dielectronic recombination rate coefficient for Fe$^{18+}+{\rm e}^- \rightarrow$ Fe$^{17+}$
is shown.  The measurement (solid curve) is from the Heidelberg storage ring (Savin  et al. 2002a\cite{sav02a})
with an experimental error assessed as $\lesssim 20\%$.  The theoretical IPIRDW results include the 
$1\rightarrow 2$, $2\rightarrow 2$ and $2\rightarrow 3$ core excitations in the
multi-configuration Breit--Pauli (MCBP, dotted curve) and Dirac--Fock data (MCDF, dashed curve) 
with $n_{\rm max}=\infty$.  
The worst deviation from experiments is $\lesssim$ 30\% (for the $3l3l'$ resonances) with a typical
uncertainty $\lesssim 20\%$ for the direct state-selective  coefficients to individual levels and for
the total dielectronic rate coefficient. There is excellent agreement between the Breit--Pauli
and Dirac--Fock results. The Breit--Pauli approach is used for our mass data production.}
 \label{fig:fig3-new} 
\end{figure}

\section{Experimental validation of dielectronic recombination data and the role of fields}
\label{experimental-validation}

The last decade has seen an enormous amount of experimental activity
in the area of dielectronic recombination. In particular, heavy-ion storage rings
coupled with electron-coolers have provided a wealth of data for
partial dielectronic recombination. (The partial here is by the
intermediate resonance state rather than the final state.) Most data
is of the `bundled-$n$' form, but some $l$-resolution is possible
for very low-lying states. The iso-electronic sequences studied range almost
exclusively from H-like through to Na-like and nuclear charges have ranged between
$Z$=2 and 92. In all cases, in the absence of external fields, there is rarely any 
significant disagreement with theory, i.e. outside of the experimental uncertainty.
A few of the more recent, typical, comparisons between experiment and the
results of IPIRDW calculations include: B\"{o}hm  et al. (2002\cite{boh02}), Savin  et al.
(2002a,b\cite{sav02a,sav02b}),
Brandau  et al. (2002\cite{bra02}).
In Fig.~\ref{fig:fig3-new}, we show representative comparisons of  dielectronic
recombination data for  O$^{5+}+{\rm e}^- \rightarrow $O$^{4+}$ and
Fe$^{18+}+{\rm e}^- \rightarrow $Fe$^{17+}$,
calculated in the IPIRDW approximation with {\sc autostructure} and which illustrates
the level of accuracy that can be expected of the theoretical data.

One major area of uncertainty is the role of external fields on dielectronic recombination,
and it is this more than anything that renders pointless efforts to compute (zero-density)
field-free data to an accuracy of better than $\simeq$20\% , say. It has long been known that
the high Rydberg states that frequently dominate the dielectronic recombination process can
be Stark-mixed by weak electric fields (Burgess \& Summers 1969\cite{bur69}), in particular the plasma
microfield (Jacobs  et al. 1976), and so increase the partial rate coefficients by factors of
2, or 3, or more,  over a wide range of $n$. Recently, the picture has been further
complicated by the discovery that magnetic fields, when crossed with an electric field,
strongly affect the electric field enhancement -- by reducing it in most cases
(see Robicheaux  et al. 1997\cite{rob97}, Bartsch  et al. 1999\cite{bar99},
Schippers  et al. 2000\cite{sch00} and B\"{o}hm  et al. 2001\cite{boh01}).
While this suppression of the electric field enhancement is
advantageous towards the use of field-free dielectronic recombination data, it is
disadvantageous in terms of trying to compute field-dependent data for plasma modelling.
Previously, it appeared that a reasonable approach would be to use the values of the plasma
microfield (which in turn depends on the plasma density) for the electric field strength
for use in the generation of field dependent (i.e. density dependent) data as input to
collisional--radiative modelling. This in itself ignored any further (e.g. external)
electric fields that might be present in the plasma environment, beyond the plasma 
microfield. The recognition of the importance of magnetic fields as well makes a 
comprehensive solution to dielectronic recombination in a plasma a distant goal and partial
data accurate to $\simeq$20\% as meaningful as necessary. Furthermore,  field enhancement is
sensitive to interacting resonances as well (see Robicheaux  et al. 1998\cite{rob98}) unlike the field-free
case. We do note again that high Rydberg states in a finite density plasma are brought into
LTE by (electron) collisions. A preliminary study by Badnell  et al. (1993\cite{bad93}) showed that the
effect of the plasma microfield on the density-dependent effective recombination rate
coefficient was suppressed by collisions driving high Rydberg states into LTE -- larger
values for the microfield, which lead to larger enhancements of the zero-density rate
coefficient, corresponds to denser plasmas for which collisions drive more states into LTE.

\begin{figure}[!htp]
\centering
\ \psfig{file=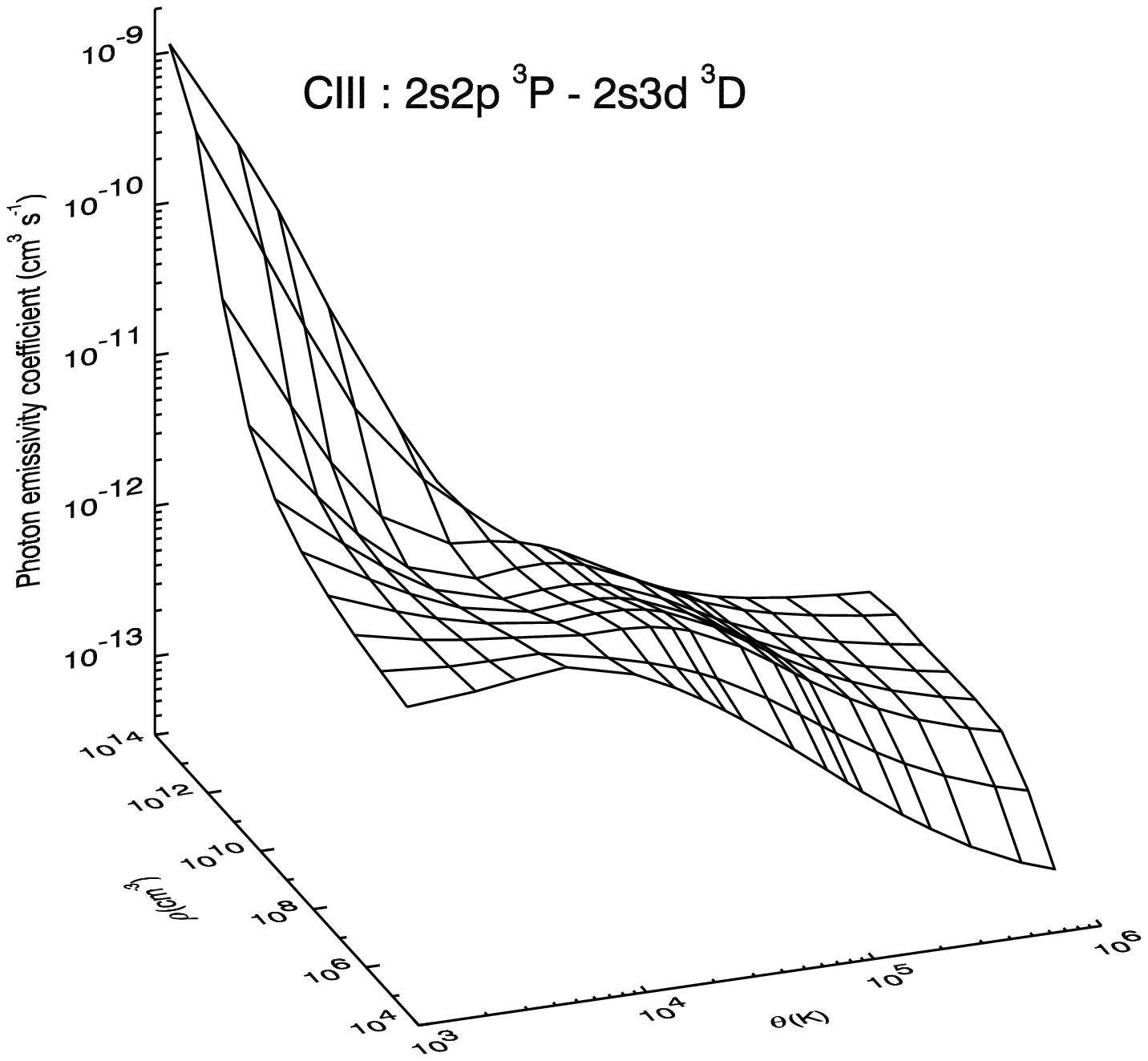,width=8.5cm}
\ \psfig{file=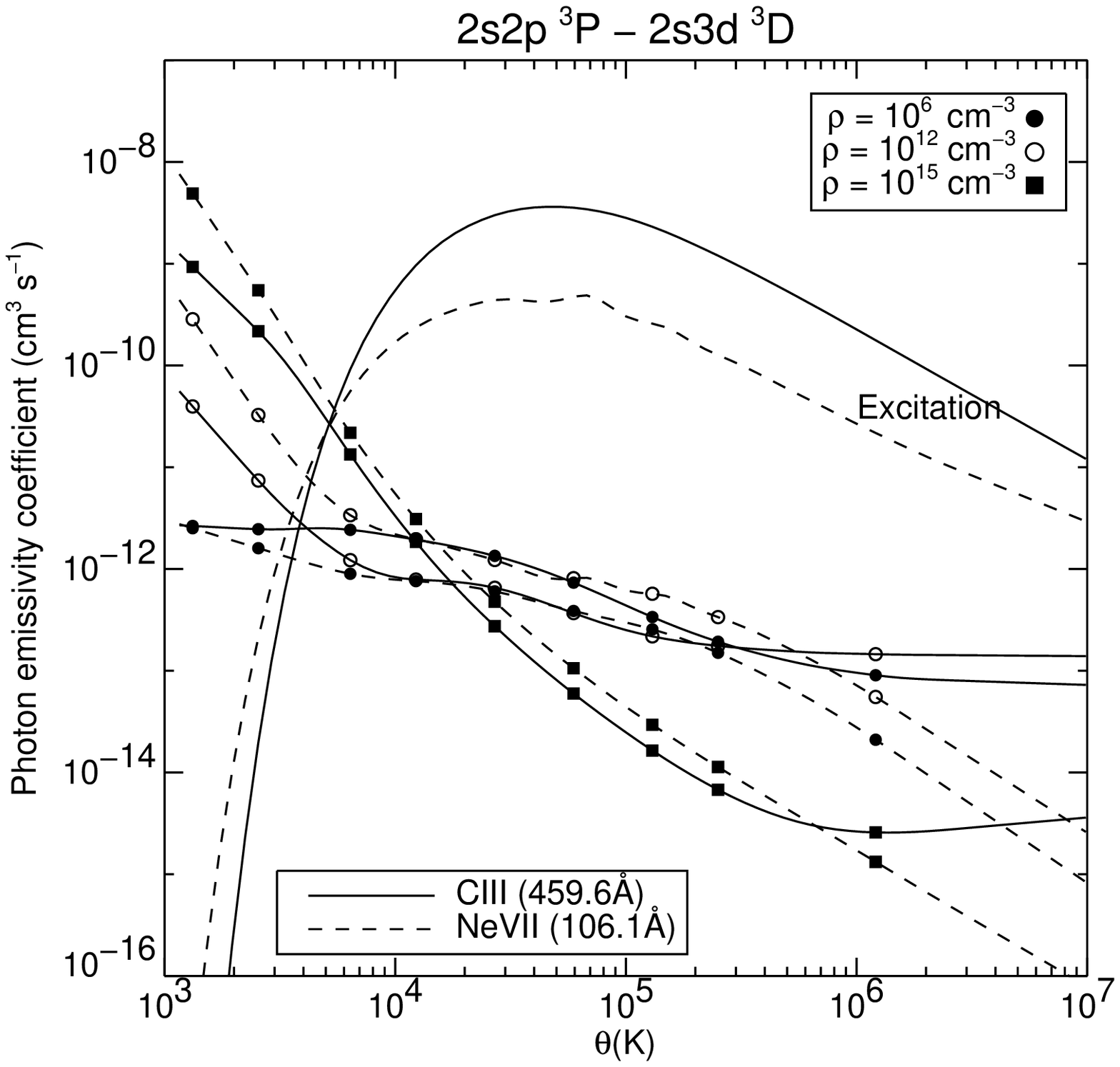,width=8.5cm}
 \caption{(a)  Temperature and density dependence of the GCR $^{\rm R}{\scriptnew
{PEC}}^{z}$ for the C~{\sc iii} $2{\rm s}2{\rm p}~^3{\rm P} - 2{\rm s}3{\rm d}~^3{\rm D}$ multiplet at 459.6~{\AA} driven by
the C$^{3+}(2{\rm s}~^2{\rm S})$ recombining ion. (b) $^{\rm R}{\scriptnew {PEC}}^{z}$ for the C~{\sc iii}
$2{\rm s}2{\rm p}~^3{\rm P} - 2{\rm s}3{\rm d}~^3{\rm D}$ multiplet at 459.6~{\AA} and Ne~{\sc vii}  $2{\rm s}2{\rm p}~^3{\rm P} -
2{\rm s}3{\rm d}~^3{\rm D}$ multiplet at 106.1~{\AA}, respectively.   Additional curves contrast the corresponding
$^{\rm X}{\scriptnew {PEC}}^{z}$s driven by the ground states of the recombined ions at  $\rho=$
10$^{12}$~cm$^{-3}$.  For comparison between  isoelectronic systems, it is convenient to use the scaled 
electron temperature $\theta=T_{\rm e}/z_1^2$ and scaled electron density $\rho=N_{\rm e}/z_1^7$, where $z_1$ is
the recombining ion charge (=3 and 7, respectively). Since the upper level is in the excited $n=3$ shell,
the cascading (projected) influence of higher levels is larger than for spectrum lines originating in the
$n=2$ shell.  The low temperature behaviour is that of the radiative recombination  process at low density,
but rises to that of the collective three-body process at high density.}
 \label{fig:fig4-new} 
\end{figure}
\begin{figure}[!htp]
\centering
\ \psfig{file=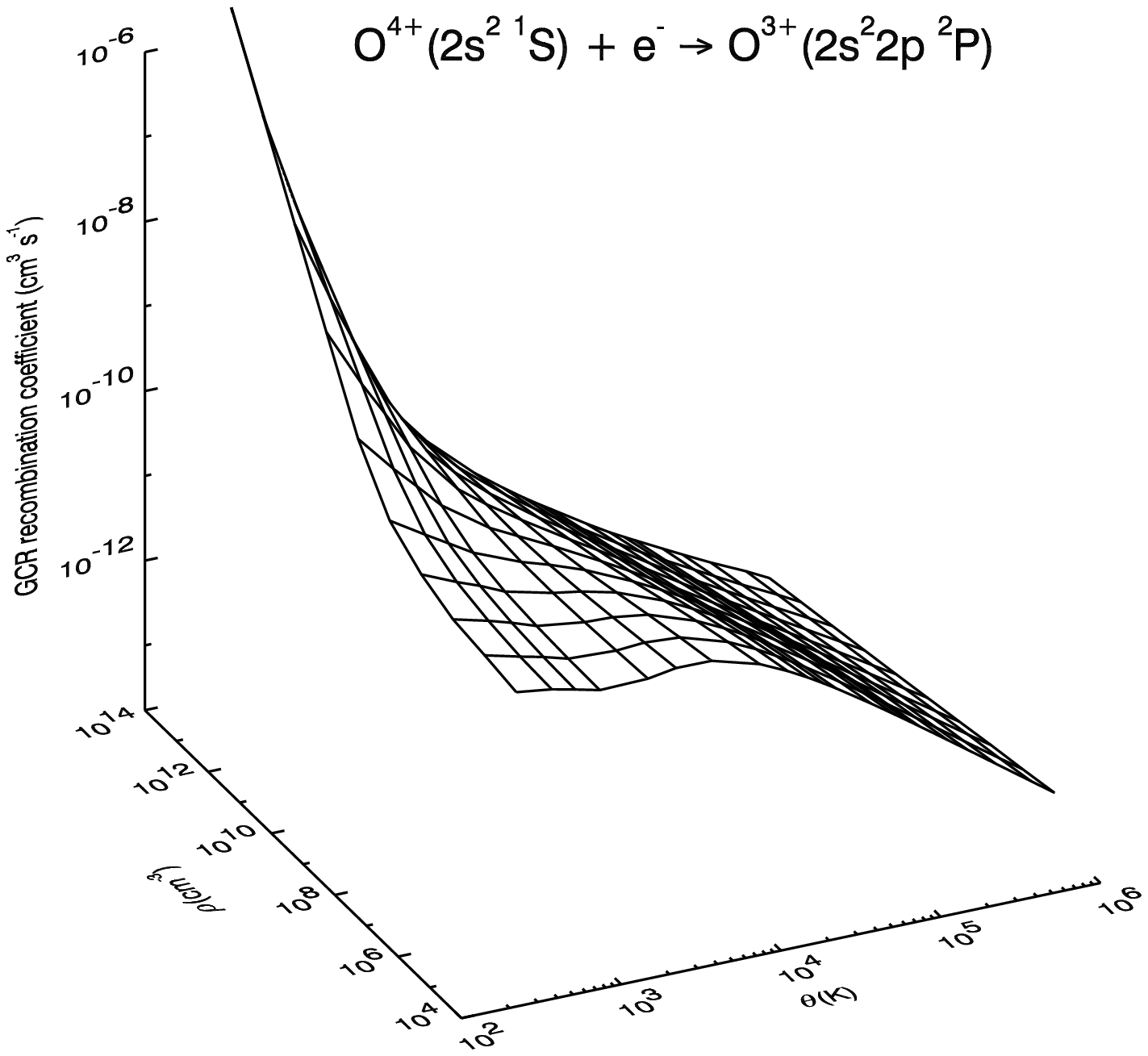,width=8.5cm}
\ \psfig{file=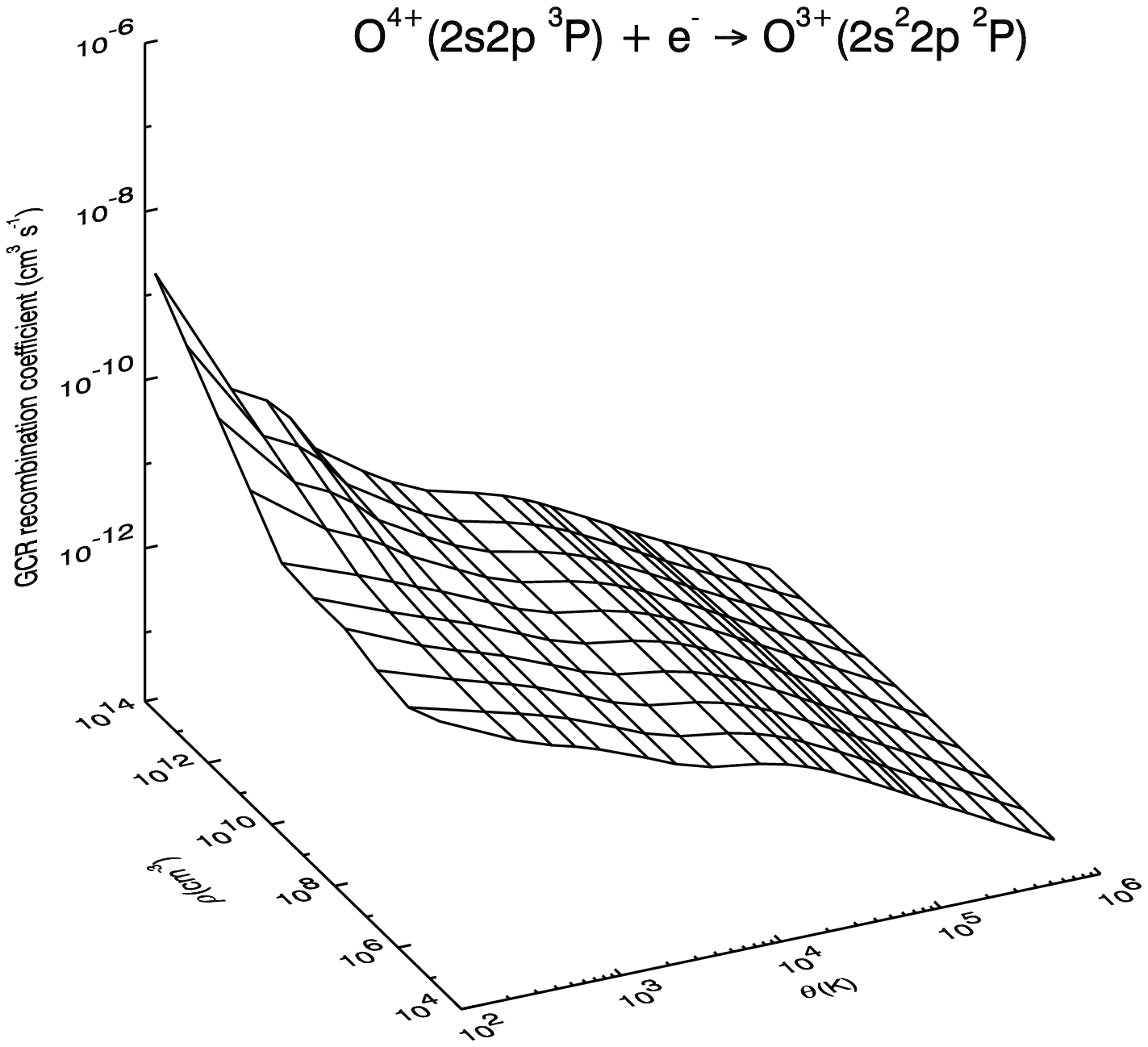,width=8.5cm}
 \caption{(a) O$^{4+}+{\rm e}^- \rightarrow $O$^{3+}$ term-resolved GCR recombination
coefficients that are driven from the $2{\rm s}^2~^1{\rm S}$ ground metastable of the O$^{4+}$ ion as a function
of electron temperature for a number of electron densities.  The scaled electron temperature, $\theta$, and
scaled electron density, $\rho$, are used. (b) O$^{4+}+{\rm e}^- \rightarrow $O$^{3+}$ term-resolved GCR
recombination coefficient driven from the $2{\rm s}2{\rm p}~^3{\rm P}$ metastable of O$^{4+}$. The suppression of
the coefficients compared with those of (a) is due to spin-breakdown alternative Auger channels.  These must
be included even in term-resolved GCR modelling.}
 \label{fig:fig5-new} 
\end{figure}

\section{Comparisons and assessments of the derived data}
\label{derived-data-validation}

As described in section \ref{gcr-modelling}, the emissivity and generalized collisional--radiative
(GCR) coefficients depend {\it inter alia} on the fundamental dielectronic cross section data.  There are
two issues of concern in assessing these derived theoretical data, viz. the relative contributions
to the effective coefficients coming from many different direct and indirect pathways and, secondly,
estimation of the uncertainty in the theoretical data, which may be treated as a `working error' in
the interpretation of spectral observations from plasmas.  

For the effective photon emissivity coefficients ($\scriptnew {PEC}$s), it is firstly to be noted that 
the relative importance of the contribution from excitation ($^{\rm X}{\scriptnew {PEC}}$)
and recombination ($^{\rm R}{\scriptnew {PEC}}$) is directly proportional to the
ionization balance fractional abundances of the (metastable) `driver' populations.  The recombination
part is most significant in  transiently recombining plasmas and it is on this part only that we
focus here.  The partitioning of the collisional--radiative matrix described in section
\ref{gcr-modelling} allows us to contrast the direct capture, capture coming via the complete set of
resolved low-levels and capture via the high bundled-$n$ quantum shells, which are treated by projection.  
The relative contributions depend differentially on density since the projection part is suppressed
selectively at higher densities.  Also, electron temperature and the recombining ion charge
influence the relative importance of the dielectronic and radiative recombination contributions and
the role of the more highly-excited levels. In Fig.~\ref{fig:fig4-new}, we show the main effects with
some illustrative results from ADAS for the C~{\sc iii} $2{\rm s}2{\rm p}~^3{\rm P} - 2{\rm s}3{\rm d}~^3{\rm D}$
multiplet at 459.6~{\AA} and the Ne~{\sc vii} $2{\rm s}2{\rm p}~^3{\rm P} - 2{\rm s}3{\rm d}~^3{\rm D}$ multiplet
at 106.1~{\AA}.

Fig.~\ref{fig:fig5-new} illustrates the main features of the generalized collisional--radiative recombination
coefficients for O$^{4+}+{\rm e}^- \rightarrow $O$^{3+}$.  In the GCR term-resolved picture for light elements,
there are four coefficients associated with the pairings of the $2{\rm s}^2~^1{\rm S}$ \& $2{\rm s}2{\rm p}~^1{\rm S}$  and
$2{\rm s}^22{\rm p}~^2{\rm P}$ \& $2{\rm s}2{\rm p}^2~^4{\rm P}$ metastable terms in the recombining and recombined systems,
respectively.  As the
radiative and three-body processes are included, the low temperature and high density behaviours reflect
these contributions.  The finite-density suppression of the coefficient for the ground parent case and
the effect of alterative Auger channels are both pronounced and also depend on the ion charge.  These effects
require GCR modelling.  The simpler stage-to-stage picture introduces a significant and, generally,
unquantifiable error.  It is to be noted that the intermediate-coupled dielectronic recombination data of this project also
sustains production of fine-structure-resolved metastable GCR coefficients appropriate to medium- and 
heavy-weight elements.     

\begin{figure}[!htp]
\centering
\ \psfig{file=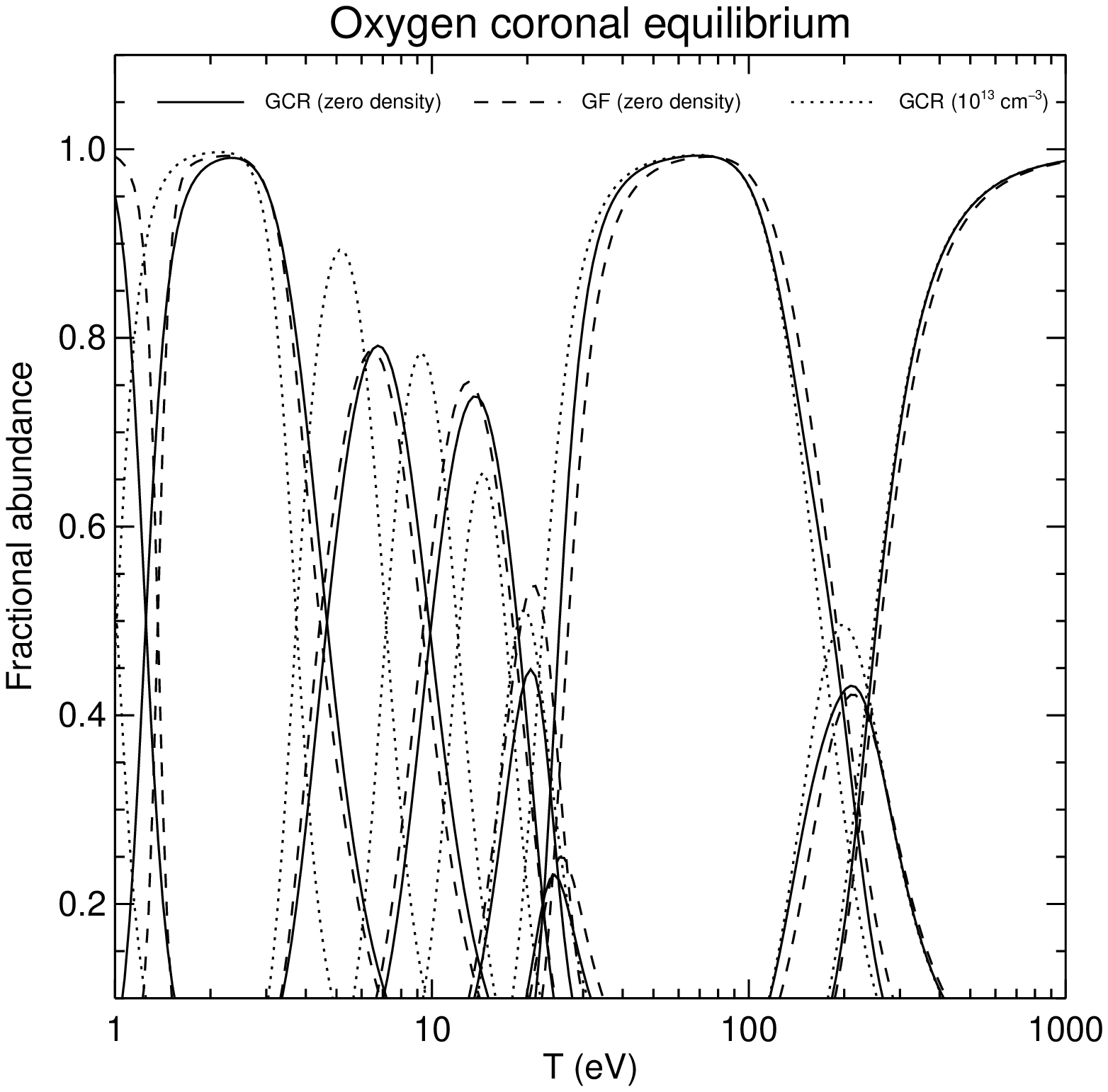,width=8.5cm}
\ \psfig{file=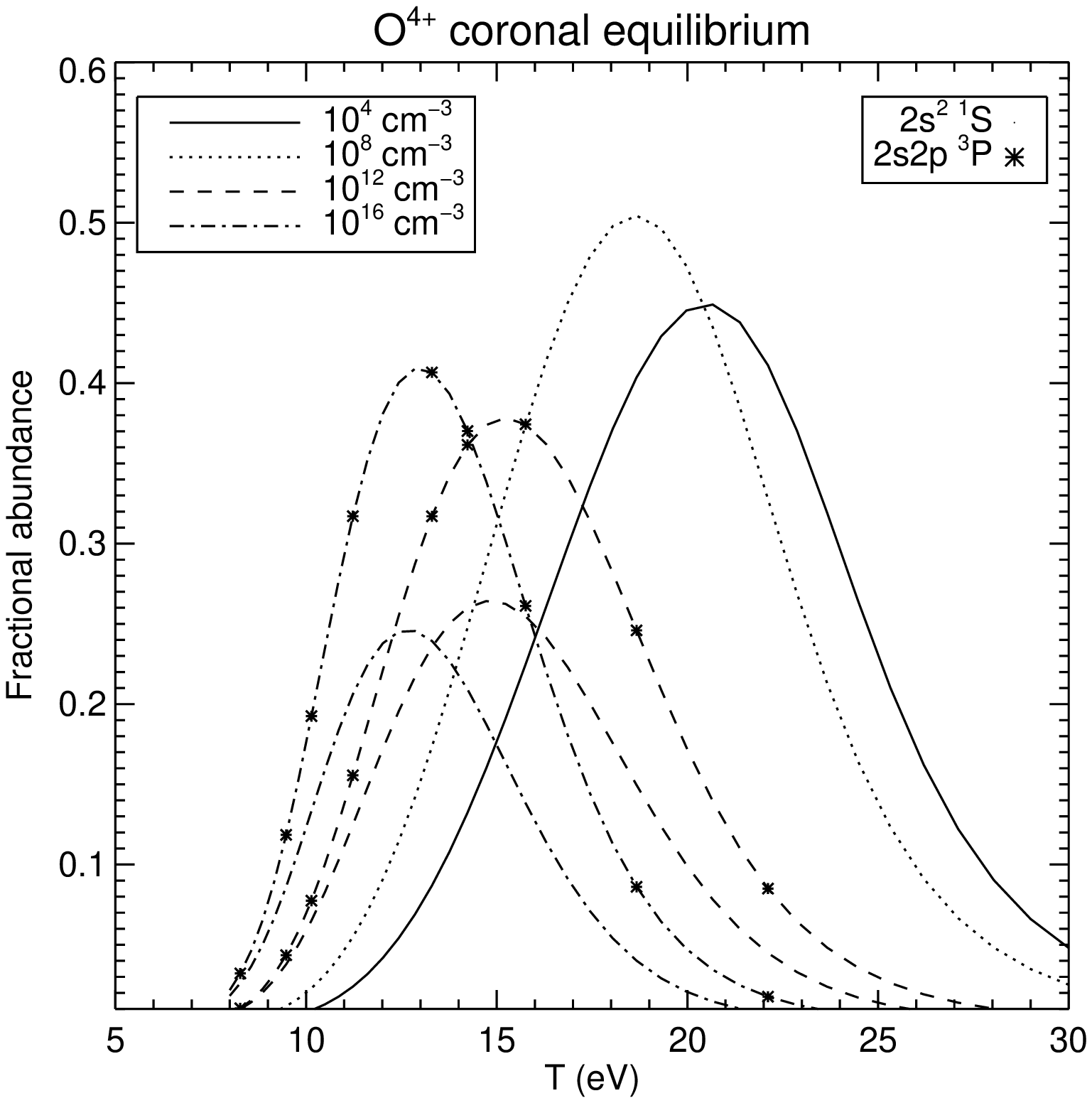,width=8.5cm}
 \caption{(a) Behaviour of the ionization balance fractional abundances for
oxygen as a function of electron temperature and density. The balance is also shown using the GF for the
dielectronic contributions. The same effective ionization rate coefficients were used in all three curve
sets, but are not considered further here.  The (level 1) GCR results are shown as a stage-to-stage balance, but
originating from a true GCR metastable-resolved calculation.  The metastable fractions are combined by
weighting with their equilibrium fractions as determined by a low-level population balance.  Note the
potential confusion between differences due to the use of a low precision zero-density dielectronic calculation, 
such as the GF, and those due to finite-density effects. 
A more complete and sophisticated approach to such metastables (extended also to ionization stages) is
called `flexible partitioning' and will be the subject of a separate work.  (b) Beryllium-like
ionization stage, O$^{4+}$, fractional abundances in the metastable term-resolved GCR picture.  Curves are
shown, therefore, for both the $2{\rm s}^2~^1{\rm S}$  and $2{\rm s}2{\rm p}~^3{\rm P}$ ground and metastable terms.
(The metastable curves are completely supressed at the lowest two densities.)
}
 \label{fig:fig6-new} 
\end{figure}

The present paper's main concern is with ensuring the quality and completeness of dielectronic data
for plasma modelling and not with all the consequential modelling of populations and ion
distributions in plasmas.  There are, however, two points to draw attention to.  Firstly, it is well
known that dielectronic recombination shifts equilibrium ionization balance fraction curves to
higher temperatures.  This is most pronounced for the ions with one or two electrons outside of closed
shells and produces a characteristic `piling-up' of these stages.  It is also these ion fractions
which show most markedly the effect of finite density  reduction of the collisional--radiative
coefficients. This cannot be ignored for moderately ionized systems in plasmas with $N_{\rm e}\gtrsim
$10$^{10}$~cm$^{-3}$.  These effects are shown in Fig.~\ref{fig:fig6-new} for oxygen.  Secondly, most
plasma transport models work only with whole ionization stage populations.  The present work,
however, sustains the metastable-resolved GCR picture.  GCR coefficients and ionization balance
fractional abundances must be bundled back to the ionization stage for such models, at the expense
of precision. Fig.~\ref{fig:fig6-new}b illustrates the resolved picture for the beryllium-like
ionization stage of oxygen.  The simplest bundling strategy imposes equilibrium fractions on the
metastable populations relative to the ground, as is used for the stage-to-stage fractional
abundances in Fig.~\ref{fig:fig6-new}a.
\begin{figure}[!htp]
\centering
\ \psfig{file=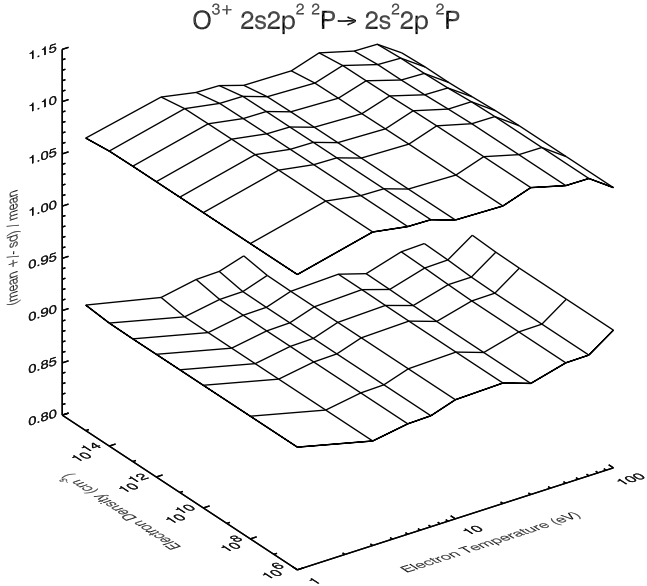,width=8.5cm}
 \caption{Upper and lower cumulative statistical ($\pm 1$ standard deviation)   surfaces  for the  
$^{\rm R}{\scriptnew {PEC}}^{z}$ for the O~{\sc iv} $2{\rm s}2{\rm p}^2~^2{\rm P} - 2{\rm s}^2 2{\rm p}~^2{\rm P}$ 
multiplet at
554.4~{\AA}. Note that projection has a substantial influence on the line emissivity and so its influence on the
cumulative propagated error is significant.  Projection and its error contribution is not included in the graph
shown. ${\scriptnew {PEC}}^{z}$ data are archived in ADAS data format {\it adf15}.  The propagated `locked
parameter' errors  are available in {\it .err} files paralleling the  {\it .dat} files for use in plasma
modelling.  }
 \label{fig:fig7-new} 
\end{figure}

Modern good practice requires an estimate of uncertainty in derived theoretical data so that meaningful deductions
can be drawn from the comparison with observations.  It is unfortunately the case that most theoretical
dielectronic data has no error associated with it.  Because of the relative complexity of dielectronic
recombination and the many contributions, agreements between different theories and with observations sometimes
appear fortuitously and do not reflect the underlying reliability (see Savin  et al. 2002a\cite{sav02a}). For the present derived
GCR coefficients and $^{\rm R}{\scriptnew {PEC}}$, we outline our approach to procuring a relevant `working error'.

In the ADAS project (Summers  et al. 2002\cite{sum02}), a distinction is made between `locked' parameters, as distinct from
`search' parameters, in the optimized fitting of models to observations.  Search parameters return a fit
uncertainty or confidence level, the locked parameters must carry an error with them.  An effective rate
coefficient is such a locked parameter.  Its uncertainty, called the cumulative statistical error, is computed from
the errors of the fundamental reaction rates as follows: Monte Carlo samples are made of all the individual
reaction coefficients, within their (assumed) independent Gaussian uncertainty distributions, and the derived
coefficient calculated.  The process is repeated many times until statistics are built up. The accumulated results
are fitted with a Gaussian variance. 

The key issue then is the starting point of uncertainties in the fundamental component dielectronic
coefficients. The BBGP codes described in section \ref{burgess-bethe} have been arranged to generate {\it adf09}
{\it baseline} files.  Such a file is differenced with the matching {\it level 1}  file to provide an error estimate for
the {\it baseline} values and may be stored in a {\it .err} file exactly paralleling the naming of the actual {\it .dat} file. 
In like manner, the {\it level 1} file may be differenced with the {\it level 2} file (averaged-over fine-structure) to
provide the {\it .err} file for {\it level 1}.   We treat this also as a conservative error for {\it level 2}.  It is
emphasized that this is not a confident absolute error, but a (hopefully) helpful appraisal of the theoretical
data.  It is most appropriate for the $n$- and $nl$- shell bundled data.  The experimental comparisons of the type
discussed in section \ref{experimental-validation}  indicate that a minimum uncertainty $\simeq$20\% is appropriate
for the term and level selective dielectronic data. Use of such {\it .err} files is not yet a common practice and its
handling within a projection matrix framework is complex.   $^{\rm R}{\scriptnew {PEC}}^{z}$ error surfaces are shown
in Fig.~\ref{fig:fig7-new} using the ADAS procedure, but propagating error only from the state-selective part. The
full handling of error will be treated in a separate paper.                

\begin{figure}[!htp]
\centering
\ \psfig{file=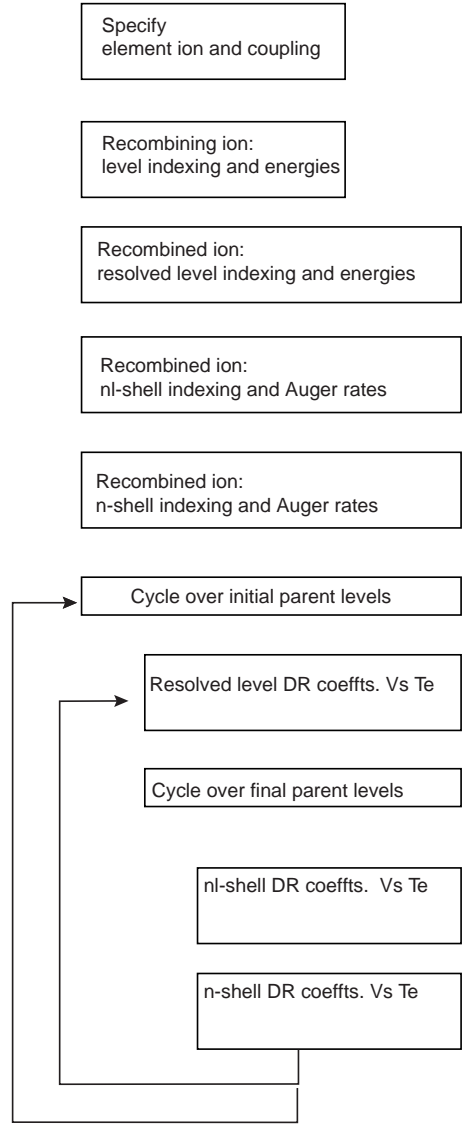,width=6.0cm}
 \caption{Organization of data within the {\it adf09} format.  {\it adf09} specifies
both LS- and intermediate-coupled data organizations.  In the LS-coupled case, the coefficients span resolved
terms with valence electron up to $n=7$; $\approx 40$ representative $n$-shells up to $n=999$.  In the intermediate-coupled
case, the coefficients span resolved levels with valence electron up to $n=7$; all $nl$-shells up to
$n=10$; $\approx 40$ representative $n$-shells up to $n=999$. The coefficients are tabulated at 19 scaled
temperatures spanning from $10$ -- $10^7$~K.  Auger rates for ionization to alternate metastable
parents for the set of $nl$-shell and $n$-shell spectators built on each parent-metastable are included for
model completeness.  The detailed specification is in Appendix A of the ADAS User's Manual (Summers 2001).}
 \label{fig:fig8-new} 
\end{figure}
\begin{figure}[!htp]
\centering
\ \psfig{file=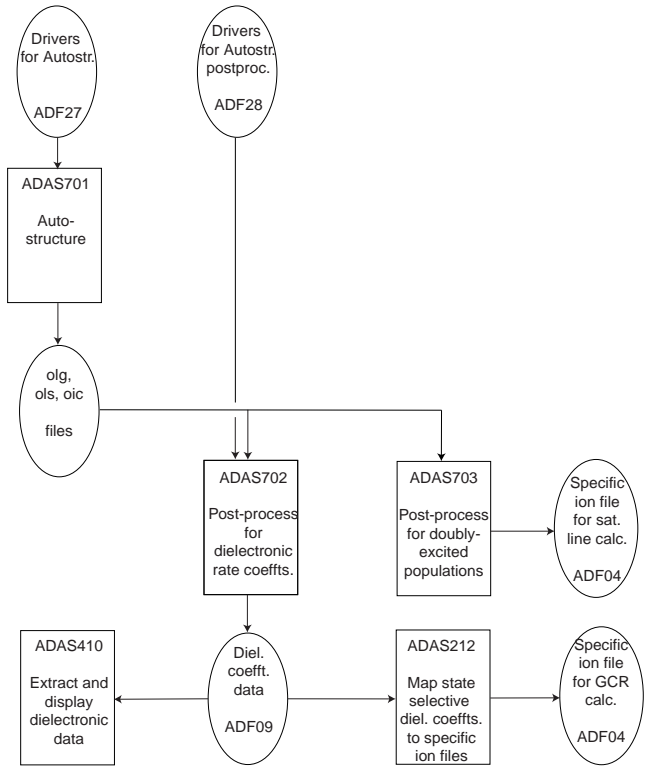,width=8.5cm}
 \caption{ADAS code organization for
production of the complete set of dielectronic data.  ADAS701 is the {\sc autostructure} code.  The {\sc
adasdr} post-processor code, ADAS702, prepares the level-resolved, bundled-$nl$ and bundled-$n$ partial
recombination coefficient data according to the {\it adf09} format specification (Summers 2001).  The code
ADAS212 maps the level-resolved data onto the specific ion file class {\it adf04}.  ADAS703 is an additional
post-processor for dielectronic satellite line modelling  and will be the subject of a separate paper.  In a
separate code chain (not shown here), ADAS807 prepares cross-referencing files to the bundled-$nl$ and
bundled-$n$ data which are required for the very high-level population calculations and the evaluation of
the projection matrices by ADAS204.  The fully-configured {\it adf04} files, together with the projection
matrices, are processed by ADAS208 which delivers the final generalized collisional--radiative (GCR) coefficients
and effective emission coefficients.}
 \label{fig:fig9-new} 
\end{figure}

\section{Structure and access to the database}
\label{database}
The complete set of dielectronic recombination data (both `partial' and `total')
will be publically available as {\it adf09} files
from the Controlled Fusion Atomic Data Center at the Oak Ridge National Laboratory, USA
(http://www-cfadc.phys.ornl.gov/).  These data
files are simple ascii text in a formatted organization. The layout differs slightly between
the {\it level 1} LS-coupled and {\it level 2} intermediate-coupled forms.  Fig.~\ref{fig:fig8-new}
summarizes the intermediate-coupled form. The document {\it appxa\_09.pdf} of Appendix A of the ADAS user
manual (available at: http://adas.phys.strath.ac.uk) provides the detailed description.  This
includes a summary of the sublibraries and their current status, content of the data lines and
the meanings of all parameters, together with some samples of the format. An example of a
pathway to a member is {\it /../adf09/jc00\#li/jc00\#li\_ne7ic23.dat} which distinguishes the producer
initials `$jc$', year number `00', recombining ion iso-electronic sequence `$li$', element `$ne$',
coupling `$ic$' and parent $n=2 \rightarrow n=3$ transition group `23'. 

We have found it convenient to archive also the driver data sets, which
initiate the {\it level 1} and {\it level 2} calculations, as other ADAS data
formats ({\it adf27} and {\it adf28}).  These have pathways which parallel {\it
adf09}. Various ADAS codes execute the primary calculations and the subsequent
collisional--radiative modelling.  The flow of calculation is summarized in
Fig.~\ref{fig:fig9-new}.

\section{Summary}

We have described the goals and methodology behind a programme to calculate a
comprehensive database of dielectronic recombination data for the 
collisional--radiative modelling of dynamic finite-density plasmas and
illustrated its use in such environments.
The first phase of the program covering H- through Ne-like sequences is
under way and illustrative results, comparisons, and total (zero-density)
rates will be the subject of a series of papers to be submitted to A\&A
in the near future, e.g. O-like ions, Zatsarinny  et al. (2003\cite{zat03}).

\end{document}